\documentclass[onefignum,onetabnum]{siamart190516}

\usepackage{amsmath}
\usepackage[noend]{algpseudocode}
\usepackage{algorithm}

\usepackage{array}
\usepackage{amsfonts}
\usepackage{graphicx}
\usepackage{epstopdf}
\usepackage{amssymb}
\usepackage{tikz,pgfplots}
\usetikzlibrary{snakes,arrows,shapes,trees}
\usepackage[position=top,aboveskip=5pt,labelformat=empty]{subfig}
\usepackage{xcolor, graphicx}
\usepackage{listings}
\usepackage{adjustbox}
\usepackage{longtable}
\usepackage{multirow}
\usetikzlibrary{calc}
\usepackage{amssymb}
\usepackage{pgfplots}
\usepackage{pgfplotstable}
\usetikzlibrary{arrows,shapes,plotmarks}
\pgfplotsset{compat=1.8}
\usepackage{soul}
\usetikzlibrary{patterns}

\ifpdf
\DeclareGraphicsExtensions{.eps,.pdf,.png,.jpg}
\else
\DeclareGraphicsExtensions{.eps}
\fi

\newsiamremark{remark}{Remark}
\newsiamremark{hypothesis}{Hypothesis}
\crefname{hypothesis}{Hypothesis}{Hypotheses}
\newsiamthm{claim}{Claim}

\headers{Scalable Local Timestepping on Octree Grids}{Milinda Fernando, Hari Sundar}

\title{Scalable Local Timestepping on Octree Grids\thanks{Submitted to the editors August 17, 2020.
\funding{This work was supported by the National Science Foundation(NSF) grant PHY-1912930, and the  National Aeronautics and Space Administration (NASA) grant 80NSSC20K0528. This work used computing resources of the Extreme Science and Engineering Discovery Environment (XSEDE) allocation TG-PHY180054 and Frontera pathway allocation PHY20033.}}}

\author{Milinda Fernando \thanks{School of Computing, University of Utah. 
  (\email{milinda@cs.utah.edu}).}
\and Hari Sundar\thanks{School of Computing, University of Utah.  
  (\email{hari@cs.utah.edu}).}}

\definecolor{bgblue}{RGB}{245,243,253}

\newcommand{\oTo}{\textsc{o2o}}
\newcommand{\oTn}{\textsc{o2n}}

\newcommand{\subsubsubsection}[1]{\paragraph{#1}\mbox{}}
\pgfplotsset{compat=1.13}

\newcommand{\dgn}{\textit{octant local nodes}}
\newcommand{\cgn}{\textit{shared octant nodes}}
\newcommand{\unzip}{\textit{unzip}}
\newcommand{\zip}{\textit{zip}}

\newcommand{\ziped}{\textit{zipped}}

\newcommand{\nuts}{\textsc{LTS}}
\newcommand{\lts}{\textsc{LTS}}
\newcommand{\uts}{\textsc{GTS}}
\newcommand{\gts}{\textsc{GTS}}

\newcommand{\maxDepth}{\textsc{maxdepth}}

\newcommand{\Frontera}{\href{https://frontera-portal.tacc.utexas.edu/}{Frontera}}

\usetikzlibrary{shapes,arrows,decorations.markings,trees,snakes}
\usepackage{accents}

\definecolor{skyblue1}{rgb}{0.447,0.624,0.812}
\definecolor{scarletred1}{rgb}{0.937,0.161,0.161}

\newdimen\HilbertLastX
\newdimen\HilbertLastY
\newcounter{HilbertOrder}

\def\DrawToNext#1#2{%
	\advance \HilbertLastX by #1
	\advance \HilbertLastY by #2
	\pgfpathlineto{\pgfqpoint{\HilbertLastX}{\HilbertLastY}}
}

\def\Hilbert[#1,#2,#3,#4,#5,#6,#7,#8] {
	\ifnum\value{HilbertOrder} > 0%
	\addtocounter{HilbertOrder}{-1}
	\Hilbert[#5,#6,#7,#8,#1,#2,#3,#4]
	\DrawToNext {#1} {#2}
	\Hilbert[#1,#2,#3,#4,#5,#6,#7,#8]
	\DrawToNext {#5} {#6}
	\Hilbert[#1,#2,#3,#4,#5,#6,#7,#8]
	\DrawToNext {#3} {#4}
	\Hilbert[#7,#8,#5,#6,#3,#4,#1,#2]
	\addtocounter{HilbertOrder}{1}
	\fi
}

\def\hilbert((#1,#2),#3){%
	\advance \HilbertLastX by #1
	\advance \HilbertLastY by #2
	\pgfpathmoveto{\pgfqpoint{\HilbertLastX}{\HilbertLastY}}
	\setcounter{HilbertOrder}{#3}
	\Hilbert[1mm,0mm,-1mm,0mm,0mm,1mm,0mm,-1mm]
	\pgfusepath{stroke}%
}

\ifpdf
\DeclareGraphicsExtensions{.eps,.pdf,.png,.jpg}
\else
\DeclareGraphicsExtensions{.eps}
\fi

\definecolor{cpu3}{HTML}{F44336}
\definecolor{cpu4}{HTML}{2196F3}
\definecolor{cpu1}{HTML}{4CAF50}
\definecolor{cpu2}{HTML}{FFC107}
\definecolor{gpu3}{HTML}{EF9A9A}
\definecolor{gpu4}{HTML}{90CAF9}
\definecolor{gpu1}{HTML}{A5D6A7}
\definecolor{gpu2}{HTML}{FFE082}

\definecolor{cpu5}{HTML}{9932CC}

\definecolor{sq_b1}{RGB}{37,52,148}
\definecolor{sq_b2}{RGB}{44,127,184}
\definecolor{sq_b3}{RGB}{65,182,196}
\definecolor{sq_b4}{RGB}{127,205,187}
\definecolor{sq_b5}{RGB}{199,233,180}
\definecolor{sq_b5}{RGB}{255,255,204}

\definecolor{sq_r1}{RGB}{189,0,38}
\definecolor{sq_r2}{RGB}{240,59,32}
\definecolor{sq_r3}{RGB}{253,141,60}
\definecolor{sq_r4}{RGB}{254,178,76}
\definecolor{sq_r5}{RGB}{254,217,118}
\definecolor{sq_r6}{RGB}{255,255,178}

\definecolor{sq_g1}{RGB}{0,104,55}
\definecolor{sq_g2}{RGB}{49,163,84}
\definecolor{sq_g3}{RGB}{120,198,121}
\definecolor{sq_g4}{RGB}{173,221,142}
\definecolor{sq_g5}{RGB}{217,240,163}
\definecolor{sq_g6}{RGB}{255,255,204}

\definecolor{div_c1}{RGB}{230,171,2}
\definecolor{div_c2}{RGB}{102,166,30}
\definecolor{div_c3}{RGB}{231,41,138}
\definecolor{div_c4}{RGB}{117,112,179}
\definecolor{div_c5}{RGB}{217,95,2}
\definecolor{div_c6}{RGB}{27,158,119}
\definecolor{div_c7}{RGB}{215,48,39}

\definecolor{lineclr}{RGB}{0,0,0}
\definecolor{utorange}{RGB}{0,0,255}
\definecolor{utsecblue}{RGB}{255,255,0}
\definecolor{utsecgreen}{RGB}{255,0,0}
\definecolor{red!15}{RGB}{0,255,255}
\definecolor{fillclr5}{RGB}{0,255,0}
\definecolor{fillclr6}{RGB}{255,0,255}
\definecolor{fillclr7}{RGB}{255,255,255}
\definecolor{fillclr8}{RGB}{0,0,0}

\def\drawcubeI(#1,#2,#3,#4,#5){ 
	\coordinate (O) at (#1,#2,#3);
	\coordinate (A) at (#1,#2+#4,#3);
	\coordinate (B) at (#1,#2+#4,#3+#4);
	\coordinate (C) at (#1,#2,#3+#4);
	\coordinate (D) at (#1+#4,#2,#3);
	\coordinate (E) at (#1+#4,#2+#4,#3);
	\coordinate (F) at (#1+#4,#2+#4,#3+#4);
	\coordinate (G) at (#1+#4,#2,#3+#4);
	\draw[#5] (O) -- (C) -- (G) -- (D) -- cycle;
	\draw[#5] (O) -- (A) -- (E) -- (D) -- cycle;
	\draw[#5] (O) -- (A) -- (B) -- (C) -- cycle;
	\draw[#5] (D) -- (E) -- (F) -- (G) -- cycle;
	\draw[#5] (C) -- (B) -- (F) -- (G) -- cycle;
	\draw[#5] (A) -- (B) -- (F) -- (E) -- cycle;
}

\def\drawcubeII(#1,#2,#3,#4,#5,#6,#7){ 
	\coordinate (O) at (#1,#2,#3);
	\coordinate (A) at (#1,#2+#4,#3);
	\coordinate (B) at (#1,#2+#4,#3+#4);
	\coordinate (C) at (#1,#2,#3+#4);
	\coordinate (D) at (#1+#4,#2,#3);
	\coordinate (E) at (#1+#4,#2+#4,#3);
	\coordinate (F) at (#1+#4,#2+#4,#3+#4);
	\coordinate (G) at (#1+#4,#2,#3+#4);
	\draw[#5,fill=#6,opacity=#7] (O) -- (C) -- (G) -- (D) -- cycle;
	\draw[#5,fill=#6,opacity=#7] (O) -- (A) -- (E) -- (D) -- cycle;
	\draw[#5,fill=#6,opacity=#7] (O) -- (A) -- (B) -- (C) -- cycle;
	\draw[#5,fill=#6,opacity=#7] (D) -- (E) -- (F) -- (G) -- cycle;
	\draw[#5,fill=#6,opacity=#7] (C) -- (B) -- (F) -- (G) -- cycle;
	\draw[#5,fill=#6,opacity=#7] (A) -- (B) -- (F) -- (E) -- cycle;
}

\def\drawNodes(#1,#2,#3,#4,#5,#6,#7){ 
	\foreach \x in {#1,#7,...,#2}{
		\foreach \y in {#3,#7,...,#4}{
			\foreach \z in {#5,#7,...,#6}{
				\draw[fill=red!60] (\x,\y,\z) circle (0.15);
			}
		}
	}				
	
}

\makeatletter
\newcommand\resetstackedplots{
	\makeatletter
	\pgfplots@stacked@isfirstplottrue
	\makeatother
	\addplot [forget plot,draw=none] coordinates{(16,0) (32,0) (64,0) (128,0) (256,0) (512,0) (1024,0) (2048,0) (4096,0) (8192,0) (16384,0) (32768,0) (65536,0) (131072,0) (262144,0)};
}
\makeatother

\makeatletter
\newcommand\resetstackedplotsOne{
	\makeatletter
	\pgfplots@stacked@isfirstplottrue
	\makeatother
	\addplot [forget plot,draw=none] coordinates{(8,0) (16,0) (32,0) (64,0) (128,0) (256,0) (512,0) (1024,0) (2048,0) (4096,0) (8192,0) (16384,0)};
}
\makeatother

\makeatletter
\newcommand\resetstackedplotsTwo{
	\makeatletter
	\pgfplots@stacked@isfirstplottrue
	\makeatother
	\addplot [forget plot,draw=none] coordinates{(64,0) (128,0) (256,0) (512,0) (1024,0) (2048,0) (4096,0) (8192,0)};
}
\makeatother

\makeatletter
\newcommand\resetstackedplotsThree{
	\makeatletter
	\pgfplots@stacked@isfirstplottrue
	\makeatother
	\addplot [forget plot,draw=none] coordinates{(8,0) (16,0) (32,0) (64,0) (128,0) (256,0) (512,0) (1024,0) (2048,0)};
}
\makeatother

\makeatletter
\newcommand\resetstackedplotsFour{
	\makeatletter
	\pgfplots@stacked@isfirstplottrue
	\makeatother
	\addplot [forget plot,draw=none] coordinates{(4,0) (8,0) (16,0) (32,0) (64,0)};
}
\makeatother

\ifpdf
\hypersetup{
  pdftitle={Scalable Local Timestepping on Octree Grids},
  pdfauthor={Milinda Fernando, Hari Sundar}
}
\fi

\begin{document}

\maketitle

\begin{abstract}
	Numerical solutions of hyperbolic partial differential equations(PDEs) are ubiquitous in science and engineering. Method of lines is a popular approach to discretize PDEs defined in spacetime, where space and time are discretized independently. When using explicit timesteppers on adaptive grids, the use of a global timestep-size dictated by the finest grid-spacing leads to inefficiencies in the coarser regions. 
	Even though adaptive space discretizations are widely used in computational sciences, temporal adaptivity is less common due to its sophisticated nature. In this paper, we present highly scalable algorithms to enable local timestepping (\nuts) for explicit timestepping schemes on fully adaptive octrees. We demonstrate the accuracy of our methods as well as the scalability of our framework across 16K cores in TACC's \Frontera. We also present a speed up estimation model for \nuts, which predicts the speedup compared to global timestepping(\gts) with an average of $0.1$ relative error. 
\end{abstract}



\section{Introduction}
\label{sec:intro}
The numerical solution of hyperbolic Partial differential equations (PDEs) plays an important role in science in engineering, with a wide range of applications from modeling earthquakes \cite{bao1998large} to simulating gravitational waves \cite{Fernando2018_GR,EINSTEIN_TOOLKIT,had_webpage}. These show up commonly as initial value problems that are typically solved using the method of lines by first discretizing in space and then solving the resulting set of ordinary differential equations (ODEs). 
This is commonly done using timestepping schemes, with explicit timestepping methods such as the Runge-Kutta methods \cite{gottlieb1998total} being more common for evolving hyperbolic systems.
Additionally, these systems are characterized by the need for high levels of spatial adaptivity \cite{mantle, EINSTEIN_TOOLKIT,Fernando2018_GR}. High levels of adaptivity impose severe stability restrictions for the explicit timestepping schemes popular for solving such systems. Therefore, it is common---especially for large scale distributed memory codes---to use a global (everywhere in space) timestep, which is   
dictated by the smallest grid resolution in space \cite{CFL}. This is highly inefficient for large systems with several orders of magnitude difference between the finest and coarsest grid resolution, as the coarser regions are forced to take extremely small timesteps than would be needed for stability \cite{GearWells84, GROTE2013270}. Local timestepping schemes can greatly speed up such codes by ensuring that the adaptivity in space is matched by a corresponding adaptivity in time. In this work, we develop an efficient, scalable local timestepping scheme (\lts) for explicit single and multi-stage timestepping algorithms on octree-adaptive spatial grids. We demonstrate the efficacy of our scheme using linear and non-linear wave equations on up to $16K$ processes on TACC's Frontera. We also present a model to estimate the expected speedup from using our scheme with an average of $10\%$ relative error in the estimation of speedup. This provides a reliable way to determine in which cases the use of local timesteping can be beneficial.   



Our framework allows space adaptivity via the use of octrees (quadtrees in $2D$) and uses high-order finite difference (FD) methods for space discretization. 
Octree based adaptive space discretizations \cite{carrier1988fast,grauer1998adaptive,fuster2009simulation,nikitin2011numerical,sundar2012parallel}  are popular in large scale simulations because of their quasi-structured nature, allowing for efficient and scalable data-structures and algorithms. Our framework targets applications in computational relativity\cite{Fernando2018_GR} and uses a 3+1 decomposition of spacetime operators, to compute a space slice where time is constant and uses standard time integration methods (explicit) to evolve the space slice forward in time. The time integration method of choice has been the Runge-Kutta methods because of the larger stability region and the availability of low-storage versions. For the specific problem of simulating binary black hole mergers to estimate gravitational waves, the spatial grid is characterized by $12$-$19$ levels of adaptivity and by the need to evolve for extremely long times. These results are simulations that need to run for months on thousands of processes. The use of \lts~ for these simulations can provide \texttt{10-70x} speed up, which can greatly reduce the time and cost of obtaining gravitational waveforms.  

While the theoretical aspects of \lts~ have been an area of active research in recent years \cite{liu2010nonuniform, liu2014nonuniform, GroteMehlinMitkova15,ltsLoadBalance,AlmquistMehlin17},
performing \nuts~ in a distributed computing environment comes with additional challenges. A central bottleneck to scalability with \lts\ is the variability in computational loads for different regions of space based on their spatial adaptivity, as finely refined regions take exponentially more timesteps than coarser regions. This requires partitioning approaches that can account for such variable workloads. Additionally, since in our target applications, the meshes are dynamic, frequent re-partitioning is required, requiring fast partitioning algorithms that are able to adapt to the variable computational load. While the use of graph partitioning approaches \cite{ltsLoadBalance} is likely to produce superior partitions, the cost of partitioning our meshes in parallel makes the approach infeasible. The key contributions presented in the paper can be summarized as follows.
\begin{itemize}
    \item \textbf{Scalable \lts~ on octrees}: We present a scalable \lts~ framework for multi-stage explicit methods, on 2:1 balanced octree grids. While several community octree frameworks are available\cite{p4est,peano,dealii,SundarSampathAdavaniEtAl07}, to the best of our knowledge, they are limited to space adaptivity.
    \item \textbf{Load balancing in \lts}: The number of local timesteps needed to reach the coarsest time on the grid depends on the spatial adaptivity. This leads to load balancing issues in \lts. In order to resolve this, we propose an space filling curve (SFC)-based weighted partitioning scheme. Compared to traditional SFC-based partitioning, we compute the weighted length of the curve to achieve a balanced load for \lts partitions.
    \item \textbf{Low overhead of \lts~ compared to \gts}: We present both strong and weak scaling results for \lts~ and \gts~ approaches on octrees. These results demonstrate that local block synchronization in \lts~ followed by time interpolations have a lower cost compared to global block synchronization present in \gts~ approach.
    \item \textbf{Accuracy of \lts}: We conduct numerical experiments to demonstrate the correctness of the implemented \lts~ framework.  The presented numerical results demonstrate the accuracy of the \lts~ framework for both linear and non-linear problems.
    \item \textbf{\lts~ performance model}: We present an analytical performance modelto estimate the speed up of \lts~ over the \gts~ scheme. The analytical model is extended to compute a theoretical upper bound for the speed up that can be achieved for a given spatial adaptivity structure.
\end{itemize}



\par \textit{Organization of the paper:} The rest of the paper is organized as follows. In \S\ref{sec:background}, we give a brief motivation on the importance of \nuts~ and a quick overview of the existing state-of-the-art approaches in the field. In \S\ref{sec:methods}, we present the algorithms and methods developed in detail to compare its efficiency to \uts. In \S\ref{sec:results}, we discuss the experimental setup, demonstrate strong and weak scalability of our approach, and the accuracy of our scheme. In \S\ref{sec:conclusion}, we conclude with directions for future work. 










\section{Background}
\label{sec:background}
In comparison to spatial adaptivity, local time adaptivity is less frequently used by large-scale applications. 
Local timestepping requires additional corrections in mismatching regions in time using interpolations and/or extrapolations. Depending on the nature of the differential operator, these operations can lead to problems in stability\cite{gander2013techniques} of the numerical scheme.  
Recent \nuts~ methods are influenced by the split Runge-Kutta(RK) methods\cite{rice1960split}, where two ODE systems are integrated using different step sizes (one called \textit{active} with the smaller timestep and the other called \textit{latent} with the larger timestep size) on the grid. Corrections using interpolations are performed for the interface between the two grid regions. A complete numerical analysis of spacetime adaptive timestepping methods is complicated, but early work by Berger provided the first mathematical analysis for adaptive schemes for the wave equation \cite{berger1982adaptive,berger1985stability}. Algorithms presented in these papers, discuss two main approaches, interpolation based and coarse-mesh approximations. In the interpolation methods, the solution at the coarse mesh is used to interpolate the values needed for the finer mesh. In the coarse mesh approximation, the coarse mesh is used to take a pseudo timestep that is used by the finer mesh. In \cite{collino2006conservative}, the authors present methods for energy-conserving corrections in time for Maxwell's equations. There is a rich literature of LTS for discontinuous Galerkin methods with special focus on energy-conserving time correction operators \cite{sandu2009multirate,joly2005error,GROTE2013270,kanevsky2007application} which are important for complicated non-linear spacetime differential operators. As mentioned previously, to enable \nuts~ in a distributed parallel setting requires specialized partitioning methods to ensure load balance. Dynamic load balancing for adaptive mesh refinement is an active research area 
\cite{bastian1998load,Fernando:2017,zoltan,flaherty1998parallel}. Sophisticated hypergraph partitioning techniques have been used for LTS for the wave equation \cite{ltsLoadBalance}, but the cost of partitioning makes it prohibitively expensive for AMR applications requiring frequent re-meshing and therefore re-partitioning. 

AMR in space and spacetime is an area of active research area. Here we present a brief overview of AMR packages that focus on both space and spacetime.Block-structured or patch-based AMR is widely used in the  astrophysics and computational fluid dynamics (CFD) communities. In block-structured AMR, the adaptivity structure is predetermined and evolved during the simulation appropriately. Some codes support \lts~ with block AMR\cite{enzo,EINSTEIN_TOOLKIT,chombo}, primarily based on the Berger-Oliger AMR criteria\cite{BOAMR}. Berger-Oliger AMR criteria supports adaptivity in space and time, but requires strong constraints on the structure of the adaptivity, such as all grids at level $l+1$(child grids) should be entirely contained within the grids at level $l$ (parent grids) while grid at the same level may overlap. There exist other block-AMR codes\cite{flash}, which only supports space adaptivity and no adaptivity in time.

Another commonly used approach for large-scale AMR is octree-based AMR\cite{p4est,peano,dealii,SundarSampathAdavaniEtAl07}. In octree-based AMR, the adaptive grid is represented using quadtrees and octrees. Unlike block-based AMR, octree-based AMR has relaxed constraints on refinement, providing highly adaptive quasi-structured (point-structured) grids in space. To the best of our knowledge, currently available octree-AMR codes are limited only to space adaptivity with no support to enable adaptivity in time. In this paper, our main focus is to perform large scale \nuts~ on adaptive octree grids, we choose the simple time interpolation methods presented in the papers \cite{liu2010nonuniform,liu2014nonuniform} which form the mathematical basis for the methods presented in this work. Additional details and analysis can be found in these papers on the \nuts\ scheme used in this paper.

\~\\ 
\section{Methodology}
\label{sec:methods}
\subsection{$3+1$ decomposition of PDEs}
In this paper, we  focus on  differential operators defined on the traditional spacetime that is a 4d manifold (3 space + 1 time dimension). Let $\mathcal{L}$ be a differential operator (linear or non-linear), $\mathcal{L}: X \rightarrow Y$ where $X,Y$ are appropriate functional spaces which $\mathcal{L}$ acts upon. For example, when $\mathcal{L} \equiv \partial_t -\partial^2_{xx} $, we get the heat operator or when $\mathcal{L} \equiv \partial^2_{tt} -\partial^2_{xx} $ one attains a linear wave operator. Throughout this paper, we focus only on the operators $\mathcal{L}$, which can be transformed into evolution equation of the form (\ref{eq:evolve_eq}), which we refer to as $3+1$ decomposition of operator $\mathcal{L}$

\begin{equation} \label{eq:evolve_eq}
\partial_t u = F(t,u(t))
\end{equation}
where, for $T \in \mathcal{R}^+$,  $F:[0,T] \times W \rightarrow X$, and $W \subset X$.  For a given $s\in[0,T)$ and $\phi \in W$, the solution or the integral curve of $F$ with respect to $s, \text{ and } W$ is a map where the range is $C^{0}([s,T] \times W) \cap C^{1}([s,T] \times X) $ that satisfies (\ref{eq:evolve_eq}) on $[s,T]$ and $u(s)=\phi$. Analysis of the well-posed nature of these integral curves, are out of the scope of this paper, hence we assume these integral curves are well-posed, and can be computed with numerical timestepping, with the appropriate necessary stability constraints.  

\subsection{Adaptivity and parallelism in space}
While this paper is centered on adaptivity in time, it is closely related to our realization of adaptivity in space using octrees. We give a brief overview of our spatial adaptivity framework in this section. The framework is freely available via an MIT license\cite{dendro5} and additional details on our algorithms can be found in \cite{Fernando:2017,Fernando2018_GR,SundarSampathBiros08}.

For a given spatial domain $\Omega=[0,2^L]^3$ where $L$ is the maximum depth parameter, we use the octree data structure to represent spatial discretization. Octrees are widely used \cite{SampathSundarAdavaniEtAl08,BursteddeWilcoxGhattas11,peano,bonsai} in computational sciences for its simplicity, efficient data-structures, ease of partitioning,  and parallel scalability. For a given function $f : \Omega \rightarrow \mathcal{R}$, we use axis-aligned octrees to generate our space discretization. The adaptive mesh refinement (AMR) criteria can be application specific, in this paper, we focus on the wavelet based AMR schemes described in \cite{Fernando2018_GR}. To perform numerical computations on adaptive octrees, we need to have neighborhood information for elements (octants) as well as at the nodal level. 


\subsubsection{Octree partitioning} 
The problem size or the load varies significantly during octree construction, balancing, and meshing during the simulation. This necessitates the efficient partitioning of the octree. As is common, we use a Hilbert curve based partitioning scheme similar to \cite{Fernando:2017}, but with minor modifications. 

\subsubsection{Octree Construction and Refinement} 
The octree construction is based on expanding user-specified functions in accordance with the specified AMR criteria, but for the work presented in this paper, we use the wavelet transform and truncate the expansion (i.e. stopping the  refinement at that level) when the coefficients are smaller than a user-specified tolerance $\epsilon>0$. Intuitively, the wavelet coefficient  measures the failure of the field to be interpolated from the coarser level. In distributed memory, all processes start from the root and refine until at least $p^2$ octants are produced. These are equally partitioned across all processes. Subsequent refinements happen in an element-local fashion and are embarrassingly parallel. A re-partition is performed at the end of construction to ensure load balance.

\subsubsection{2:1 Balancing} 
Following the octree construction, we enforce a 2:1 balance condition, i.e., any octant can have a neighbor that is either the same size, half as big, or twice as big. This makes subsequent operations simpler without affecting the adaptive properties. Our balancing algorithm is similar to existing approaches for balancing octrees \cite{bern1999parallel,SundarSampathAdavaniEtAl07} with minor changes in the choice of data structures, and process-local balancing algorithms. 

\subsubsection{Mesh generation}
\label{sec:meshing} 
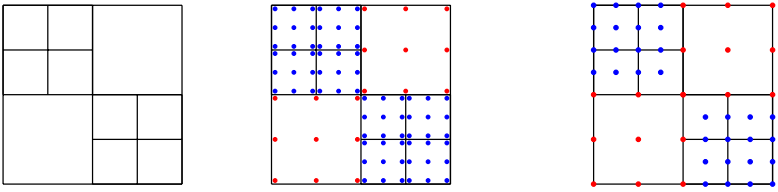
\begin{figure}
	\centering
	\resizebox{0.8\columnwidth}{!}{
	\begin{tikzpicture}[scale=0.2,every node/.style={scale=0.6}]
		
	\begin{scope}[shift={(0,0)}]
	\draw[black,step=5] (0,0) grid +(10,10);
	\draw[black,step=2.5] (5,0) grid +(5,5);
	\draw[black,step=2.5] (0,5) grid +(5,5);
	
	\end{scope}
	
	\begin{scope}[shift={(15,0)}]
	\draw[black,step=5] (0,0) grid +(10,10);
	\draw[black,step=2.5] (5,0) grid +(5,5);
	\draw[black,step=2.5] (0,5) grid +(5,5);

	\def \r{0.1}
	\foreach \x in {0.2,2.5,4.8}{
	\foreach \y in {0.2,2.5,4.8}{
		\draw[red,fill=red] (\x,\y) circle (\r);
      }
	}

	\foreach \x in {5.2,7.5,9.8}{
	\foreach \y in {5.2,7.5,9.8}{
		\draw[red,fill=red] (\x,\y) circle (\r);
      }
	}	
	
	\foreach \x in {0.2,1.25,2.3,2.7,3.75,4.8}{
	\foreach \y in {5.2,6.25,7.3}{
		\draw[blue,fill=blue] (\x,\y) circle (\r);
      }
    \foreach \y in {7.7,8.75,9.8}{
		\draw[blue,fill=blue] (\x,\y) circle (\r);
      }  
	}
	
	\foreach \x in {5.2,6.25,7.3,7.7,8.75,9.8}{
	\foreach \y in {0.2,1.25,2.3}{
		\draw[blue,fill=blue] (\x,\y) circle (\r);
      }
    \foreach \y in {2.7,3.75,4.8}{
		\draw[blue,fill=blue] (\x,\y) circle (\r);
      }  
	}
					
	\end{scope}
	
	\begin{scope}[shift={(33,0)}]

	\draw[black,step=5] (0,0) grid +(10,10);
	\draw[black,step=2.5] (5,0) grid +(5,5);
	\draw[black,step=2.5] (0,5) grid +(5,5);
	
	\def \r{0.12}
	\foreach \x in {0,2.5,5}{
	\foreach \y in {0,2.5,5}{
		\draw[red,fill=red] (\x,\y) circle (\r);
      }
	}
	
	\foreach \x in {5,7.5,10}{
	\foreach \y in {5,7.5,10}{
		\draw[red,fill=red] (\x,\y) circle (\r);
      }
	}

	\foreach \x in {0,1.25,2.5,3.75}{
	\foreach \y in {6.25,7.5,8.75,10}{
		\draw[blue,fill=blue] (\x,\y) circle (\r);
      }
	}	
	
	\foreach \x in {6.25,7.5,8.75,10}{
	\foreach \y in {0,1.25,2.5,3.75}{
		\draw[blue,fill=blue] (\x,\y) circle (\r);
      }
	}
	
	\end{scope}	
	
	\end{tikzpicture}}
\caption{\label{fig:dg_to_cg} \small A $2D$ example of \dgn~ (in the center) and \cgn~(the rightmost figure) nodal representation (for octant order of $2$) of the adaptive quadtree shown in the leftmost figure. Note that in \dgn, representation nodes are local to each octant and contain duplicate nodes. By removing all the duplicate and hanging nodes by the rule of nodal ownership, we get the \cgn~ representation. Note that the nodes are color-coded based on the octant level.}
\vspace{-0.15in}
\end{figure}

By meshing, we refer to the construction of the data structures required to perform $d^{th}$ order numerical computations octree data. 
One of the key steps of the mesh generation phase is to construct neighborhood information for octants. Primarily, there are two maps that are produced. The first  map \oTo ~to determine the neighboring octants of a given octant and a map \oTn ~to compute the nodes corresponding to a given octant. \oTo~ map is generated, by performing parallel searches  similar to approaches described in \cite{Fernando:2017, p4est} and optimized as per the methods described in \cite{Fernando2018_GR}. Assuming we have $n$ octants per partition, these search operations and building required \oTo~ and \oTn~ data structures can be performed in $\mathcal{O}(n\log(n))$ and $\mathcal{O}(n)$ complexity, respectively. Computation of the \oTn~ map is a local operation, where we start with octant local nodes( with duplicate nodes for shared octant boundaries), and eliminating duplicate nodes with a globally well defined criteria such as space filling curve (SFC) ordering operator(see figure \ref{fig:dg_to_cg}).

\subsubsection{Octree to block decomposition}
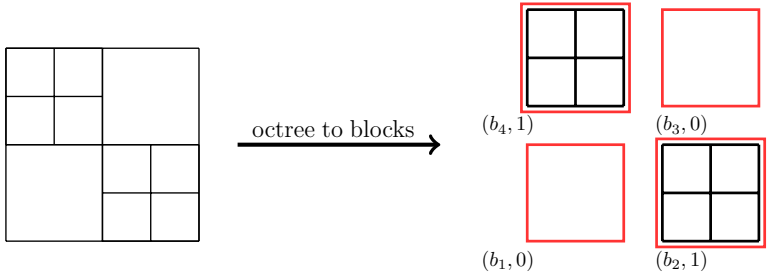
\begin{figure}[tbh]
\centering
	\resizebox{0.8\columnwidth}{!}{
	\begin{tikzpicture}[scale=0.2,every node/.style={scale=0.6}]
		
	\begin{scope}[shift={(0,0)}]
	\draw[black,step=5cm] (0,0) grid +(10,10);
	\draw[black,step=2.5cm] (5,0) grid +(5,5);
	\draw[black,step=2.5cm] (0,5) grid +(5,5);
	
	

	
	\end{scope}
	\begin{scope}[shift={(20,0)}]
	\draw[black,very thick,->,xshift=-8cm] (0,5) -- (10.5,5) node[anchor=center,above,text width=6.5cm] {\large octree to blocks};
	\end{scope}
	\begin{scope}[shift={(27,0)}]
	\draw[red!80,thick] (0,0) rectangle +(5,5);
	\draw[red!80,thick,xshift=2cm,yshift=2.0cm] (5,5) rectangle +(5,5);
	\draw[red!80,thick,xshift=0cm,yshift=2.0cm] (-0.3,4.7) rectangle +(5.6,5.6);
	\draw[red!80,thick,xshift=2cm,yshift=0.0cm] (4.7,-0.3) rectangle +(5.6,5.6);
	\draw[black,thick,step=2.5cm,xshift=2.0cm] (5,0) grid +(5,5);
	\draw[black,thick,step=2.5cm,yshift=2.0cm] (0,5) grid +(5,5);
	\draw[] (-1,-1) node{$(b_1,0)$};
	\draw[xshift=9cm] (-1,-1) node{$(b_2,1)$};
	\draw[xshift=9cm,yshift=7cm] (-1,-1) node{$(b_3,0)$};
	\draw[xshift=0cm,yshift=7cm] (-1,-1) node{$(b_4,1)$};
	\end{scope}
	
	

	
	
	
	\end{tikzpicture}
}
	
\caption{\label{fig:oct_to_blocks} \small A simplistic example of octree ($\mathcal{T}$) to block decomposition. The left figure shows the considering adaptive octree, and its block decomposition is shown in the right. Each block $(b_k,l)$ is associated with uniform grid level parameter $l$, where $l$ denotes the level of refinement of the sub-tree rooted at $b_k$ node in $\mathcal{T}$.}
\vspace{-0.15in}
\end{figure}

For a given octree $\mathcal{T}$, we compute a compressed octree of $\mathcal{T}$, denoted as $\mathcal{B}$ (blocks) where each leaf node in $\mathcal{B}$ is a node in $\mathcal{T}$ with uniformly refined sub-octree. Octree to block decomposition allows, to perform finite difference (FD) computations on adaptive octrees, as well as simplifies the computation of data structures required to perform \nuts\ (see figure \ref{fig:oct_to_blocks}). 

\subsection{Finite difference computations on adaptive octrees.}
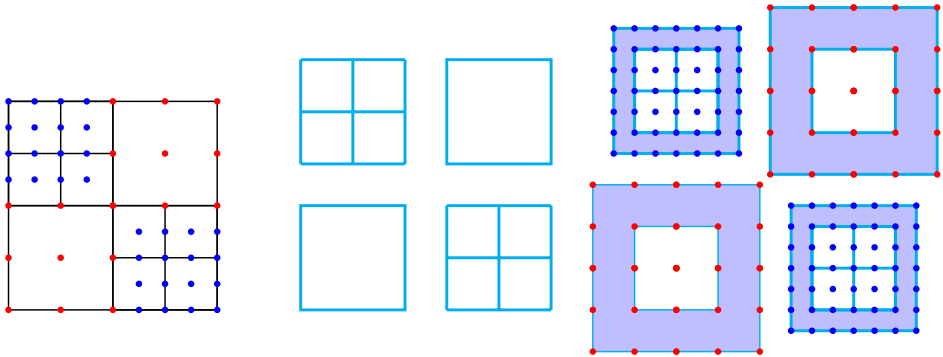
\begin{figure}[tbh]
	\resizebox{0.98\columnwidth}{!}{
	\begin{tikzpicture}[scale=0.2,every node/.style={scale=0.6}]
		
	\begin{scope}[shift={(0,0)}]
	\draw[black,step=5cm] (0,0) grid +(10,10);
	\draw[black,step=2.5cm] (5,0) grid +(5,5);
	\draw[black,step=2.5cm] (0,5) grid +(5,5);
	
	\def \r{0.12}
	\foreach \x in {0,2.5,5}{
	\foreach \y in {0,2.5,5}{
		\draw[red,fill=red] (\x,\y) circle (\r);
      }
	}
	
	\foreach \x in {5,7.5,10}{
	\foreach \y in {5,7.5,10}{
		\draw[red,fill=red] (\x,\y) circle (\r);
      }
	}

	\foreach \x in {0,1.25,2.5,3.75}{
	\foreach \y in {6.25,7.5,8.75,10}{
		\draw[blue,fill=blue] (\x,\y) circle (\r);
      }
	}	
	
	\foreach \x in {6.25,7.5,8.75,10}{
	\foreach \y in {0,1.25,2.5,3.75}{
		\draw[blue,fill=blue] (\x,\y) circle (\r);
      }
	}
	\end{scope}
	
	\begin{scope}[shift={(14,0)}]
	\draw[cyan,thick] (0,0) rectangle +(5,5);
	\draw[cyan,thick,xshift=2cm,yshift=2.0cm] (5,5) rectangle +(5,5);
	\draw[cyan,thick,step=2.5cm,xshift=2.0cm] (5,0) grid +(5,5);
	\draw[cyan,thick,step=2.5cm,yshift=2.0cm] (0,5) grid +(5,5);
	\end{scope}
	
	\begin{scope}[shift={(30,0)}]
	\def \r{0.12}
	\draw[cyan,fill=blue!50,fill opacity=0.5]
	(-2,-2)--(-2,6)--(6,6)--(6,-2)--cycle
	(0,0) -- (4,0)-- (4,4)--(0,4)--cycle;
	\foreach \x in {-2,0,2,2,4,6}{
	\foreach \y in {-2,0,2,2,4,6}{
		\draw[red,fill=red] (\x,\y) circle (\r);
      }
	}
	
	\draw[cyan,thick,fill=blue!50,fill opacity=0.5,even odd rule,xshift=8.5cm,yshift=8.5cm] 
	(-2,-2)--(-2,6)--(6,6)--(6,-2)--cycle
	(0,0) -- (4,0)-- (4,4)--(0,4)--cycle;
		\foreach \x in {-2,0,2,2,4,6}{
	\foreach \y in {-2,0,2,2,4,6}{
		\draw[red,fill=red,xshift=8.5cm,yshift=8.5cm] (\x,\y) circle (\r);
      }
	}

	\draw[cyan,thick,fill=blue!50,fill opacity=0.5,even odd rule,xshift=8.5cm] 
	(-1,-1)--(-1,5)--(5,5)--(5,-1)--cycle
	(0,0) -- (4,0)-- (4,4)--(0,4)--cycle;
	\draw[cyan,thick,step=2,xshift=8.5cm] (0,0) grid +(4,4);
	\foreach \x in {-1,0,1,2,3,4,5}{
	\foreach \y in {-1,0,1,2,3,4,5}{
		\draw[blue,fill=blue,xshift=8.5cm] (\x,\y) circle (\r);
      }
	}
	
	\draw[cyan,thick,fill=blue!50,fill opacity=0.5,even odd rule,yshift=8.5cm] 
	(-1,-1)--(-1,5)--(5,5)--(5,-1)--cycle
	(0,0) -- (4,0)-- (4,4)--(0,4)--cycle;
	\draw[cyan,thick,step=2,yshift=8.5cm] (0,0) grid +(4,4);	
	\foreach \x in {-1,0,1,2,3,4,5}{
	\foreach \y in {-1,0,1,2,3,4,5}{
		\draw[blue,fill=blue,yshift=8.5cm] (\x,\y) circle (\r);
      }
	}
	
	\end{scope}	
	\end{tikzpicture}
}	
\caption{\label{fig:unzip} \small A simplistic example of octree to block decomposition and \unzip~ operation. The leftmost figure shows the considering adaptive octree with \cgn, and its block decomposition is shown in the middle. Note that the given octree is decomposed into four regular blocks of different sizes. The rightmost figure shows the decomposed blocks padded with values coming from neighboring octants with interpolation if needed. In order to perform \unzip~ operation, both $\oTo$ and $\oTn$ mappings are used. }
\vspace{-0.15in}
\end{figure}

We use the finite difference method (FDM) to discretize differential operators in space. Performing FD computations on adaptive octrees require additional processing, to ensure that each refinement level has neighboring points available at the same resolution. This is achieved by performing octree to block decomposition where for each block, we compute a padding region of a specific width corresponding to the FD stencil. In order to compute the padded blocks, we use the computed octant to octant (\oTo) and octant to nodal (\oTn) information with the corresponding space interpolations. For example, between finer and coarser grid blocks, we use coarser to finer interpolation to compute the padding region for the finer block, while using finer to coarser injections for the padding region of the coarser block. Note that in the paper, the octant shared node representation is also referred to as \ziped~ representation (see figure \ref{fig:dg_to_cg}) and block with padding computed referred to as the {\it unzipped} representation. All the FD stencils are applied at the {\it unzipped} representation, while just prior to the communication, we perform \zip~ operation so the inter-process communication happens in the efficient more compact form (see figure \ref{fig:unzip}). 

\subsection{Explicit timestepping schemes}

Explicit timestepping (ETS) is a class of numerical schemes that compute the solution curve for (\ref{eq:evolve_eq}). In explicit methods, the solution at time $t_{n+1}$, $ u^{n+1} \equiv u^{n+1}(t_{n+1},\cdot)$ is computed directly from the solution at the previous timestep $u^{n}$ and does not require a linear solve. In order to numerically evolve (\ref{eq:evolve_eq}), we discretize $F$, i.e., discretization in space first. The resulting set of ODEs are discretized using explicit time integration. Depending on the properties on $\mathcal{L}$, there can be additional constraints on $\Delta t, \Delta x$. For most hyperbolic operators, the Courant–Friedrichs–Lewy (CFL) condition \cite{CFL} is a necessary condition for stability for numerical time evolution. The CFL condition $\frac{\Delta t}{\Delta x}< C$, where C is a constant that specifies a necessary condition for stability. Intuitively, it imposes the constraint that we cannot propagate spatial information in time, faster than the speed of information propagation defined by operator $\mathcal{L}$. Runge-Kutta (RK) \cite{nr} schemes are widely used explicit timesteppers (see equation (\ref{eq:rk_schemes})), that will be our main focus for spatially adaptive local timestepping. 

\begin{align}
   k_1 &= F(u^{n})  \nonumber \\
   k_2 &= F(u^{n} + a_{2,1}k_1 ) \nonumber  \\
   &\;\;\vdots \notag \\
   k_p &= F(u^{n} + a_{p,1}k_1 + ... + a_{p,p-1}k_{p-1} ) \nonumber  \\
   u^{n+1} &= u^{n} + \Delta t (\sum_{i=1}^{p} w_i k_i) \label{eq:rk_schemes}
\end{align}

Being one-step methods, they do not require a starting procedure, have a large stability region, and from a large-scale computational perspective can easily be converted into low-storage versions. 

\begin{figure}[tbh]
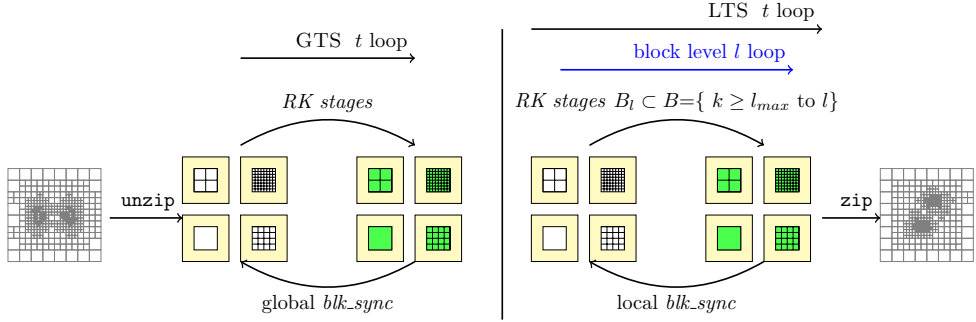

    \centering
    \resizebox{\textwidth}{!}{
    \begin{tikzpicture}
    
    \begin{scope}[xshift=-2cm,yshift=1.0cm,scale=0.025]
    \input{fig_tex/gr_init1_r1}
    \end{scope}
    \draw[thick,->] (-0.25,1.75) -- (1,1.75) node[anchor=south east] {\texttt{unzip}};
    
    \begin{scope}
    \draw[black,thick,->] (2,4.5) -- (5,4.5) node[anchor=south east] {\uts~ $t$ loop};
    \begin{scope}[xshift=1cm,yshift=1cm,scale=0.8]
    \draw[thin,fill=yellow!30] (0,0)--(0,1)--(1,1)--(1,0)--cycle
    (0.25,0.25)--(0.75,0.25)--(0.75,0.75)--(0.25,0.75)--cycle;
    \end{scope}
    \begin{scope}[xshift=1cm,yshift=2cm,scale=0.8]
    \draw[thin,fill=yellow!30] (0,0)--(0,1)--(1,1)--(1,0)--cycle
    (0.25,0.25)--(0.75,0.25)--(0.75,0.75)--(0.25,0.75)--cycle;
    \draw[black,very thin,step=0.25cm] (0.25,0.25) grid +(0.5,0.5);
    \end{scope}
    
    \begin{scope}[xshift=2.0cm,yshift=1cm,scale=0.8]
    \draw[thin,fill=yellow!30] (0,0)--(0,1)--(1,1)--(1,0)--cycle
    (0.25,0.25)--(0.75,0.25)--(0.75,0.75)--(0.25,0.75)--cycle;
    \draw[black,ultra thin,step=0.125cm] (0.25,0.25) grid +(0.5,0.5);
    \end{scope}
    \begin{scope}[xshift=2.0cm,yshift=2cm,scale=0.8]
    \draw[thin,fill=yellow!30] (0,0)--(0,1)--(1,1)--(1,0)--cycle
    (0.25,0.25)--(0.75,0.25)--(0.75,0.75)--(0.25,0.75)--cycle;
    \draw[black,ultra thin,step=0.0625cm] (0.25,0.25) grid +(0.5,0.5);
    \end{scope}
    
    \begin{scope}[xshift=4cm,yshift=1cm,scale=0.8]
    \draw[thin,fill=yellow!30] (0,0)--(0,1)--(1,1)--(1,0)--cycle
    (0.25,0.25)--(0.75,0.25)--(0.75,0.75)--(0.25,0.75)--cycle;
    \draw[black,very thin,,fill=green!70] (0.25,0.25) rectangle +(0.5,0.5);
    \end{scope}
    \begin{scope}[xshift=4cm,yshift=2cm,scale=0.8]
    \draw[thin,fill=yellow!30] (0,0)--(0,1)--(1,1)--(1,0)--cycle
    (0.25,0.25)--(0.75,0.25)--(0.75,0.75)--(0.25,0.75)--cycle;
    \draw[black,very thin,,fill=green!70,step=0.25cm] (0.25,0.25) grid +(0.5,0.5) rectangle(0.25,0.25);
    \end{scope}
    
    \begin{scope}[xshift=5cm,yshift=1cm,scale=0.8]
    \draw[thin,fill=yellow!30] (0,0)--(0,1)--(1,1)--(1,0)--cycle
    (0.25,0.25)--(0.75,0.25)--(0.75,0.75)--(0.25,0.75)--cycle;
    \draw[black,very thin,,fill=green!70,step=0.125cm] (0.25,0.25) grid +(0.5,0.5) rectangle(0.25,0.25);
    \end{scope}
    \begin{scope}[xshift=5cm,yshift=2cm,scale=0.8]
    \draw[thin,fill=yellow!30] (0,0)--(0,1)--(1,1)--(1,0)--cycle
    (0.25,0.25)--(0.75,0.25)--(0.75,0.75)--(0.25,0.75)--cycle;
    \draw[black,very thin,,fill=green!70,step=0.0625cm] (0.25,0.25) grid +(0.5,0.5)rectangle(0.25,0.25);
    \end{scope}
    
    \path [black,fill=cyan,thick,->,bend left]  (2,3.0) edge node[above] { \textit{RK stages}}(5.0,3.0);
    \path [black,fill=cyan,thick,->,bend left] (5.0,1.0) edge node[below] { global \textit{blk\_sync}} (2,1.0);
    
    \end{scope}
    
    \begin{scope}[xshift=6cm]
    \draw[black,thick,->] (1,5) -- (6,5) node[anchor=south east] {\nuts~ $t$ loop};
    \draw[thick,blue,->] (1.5,4.3) -- (5.5,4.3) node[anchor=south east] {block level $l$ loop};
    \begin{scope}[xshift=1cm,yshift=1cm,scale=0.8]
    \draw[thin,fill=yellow!30] (0,0)--(0,1)--(1,1)--(1,0)--cycle
    (0.25,0.25)--(0.75,0.25)--(0.75,0.75)--(0.25,0.75)--cycle;
    \end{scope}
    \begin{scope}[xshift=1cm,yshift=2cm,scale=0.8]
    \draw[thin,fill=yellow!30] (0,0)--(0,1)--(1,1)--(1,0)--cycle
    (0.25,0.25)--(0.75,0.25)--(0.75,0.75)--(0.25,0.75)--cycle;
    \draw[black,very thin,step=0.25cm] (0.25,0.25) grid +(0.5,0.5);
    \end{scope}
    
    \begin{scope}[xshift=2.0cm,yshift=1cm,scale=0.8]
    \draw[thin,fill=yellow!30] (0,0)--(0,1)--(1,1)--(1,0)--cycle
    (0.25,0.25)--(0.75,0.25)--(0.75,0.75)--(0.25,0.75)--cycle;
    \draw[black,ultra thin,step=0.125cm] (0.25,0.25) grid +(0.5,0.5);
    \end{scope}
    \begin{scope}[xshift=2.0cm,yshift=2cm,scale=0.8]
    \draw[thin,fill=yellow!30] (0,0)--(0,1)--(1,1)--(1,0)--cycle
    (0.25,0.25)--(0.75,0.25)--(0.75,0.75)--(0.25,0.75)--cycle;
    \draw[black,ultra thin,step=0.0625cm] (0.25,0.25) grid +(0.5,0.5);
    \end{scope}
    
    \begin{scope}[xshift=4cm,yshift=1cm,scale=0.8]
    \draw[thin,fill=yellow!30] (0,0)--(0,1)--(1,1)--(1,0)--cycle
    (0.25,0.25)--(0.75,0.25)--(0.75,0.75)--(0.25,0.75)--cycle;
    \draw[black,very thin,,fill=green!70] (0.25,0.25) rectangle +(0.5,0.5);
    \end{scope}
    \begin{scope}[xshift=4cm,yshift=2cm,scale=0.8]
    \draw[thin,fill=yellow!30] (0,0)--(0,1)--(1,1)--(1,0)--cycle
    (0.25,0.25)--(0.75,0.25)--(0.75,0.75)--(0.25,0.75)--cycle;
    \draw[black,very thin,,fill=green!70,step=0.25cm] (0.25,0.25) grid +(0.5,0.5) rectangle(0.25,0.25);
    \end{scope}
    
    \begin{scope}[xshift=5cm,yshift=1cm,scale=0.8]
    \draw[thin,fill=yellow!30] (0,0)--(0,1)--(1,1)--(1,0)--cycle
    (0.25,0.25)--(0.75,0.25)--(0.75,0.75)--(0.25,0.75)--cycle;
    \draw[black,very thin,,fill=green!70,step=0.125cm] (0.25,0.25) grid +(0.5,0.5) rectangle(0.25,0.25);
    \end{scope}
    \begin{scope}[xshift=5cm,yshift=2cm,scale=0.8]
    \draw[thin,fill=yellow!30] (0,0)--(0,1)--(1,1)--(1,0)--cycle
    (0.25,0.25)--(0.75,0.25)--(0.75,0.75)--(0.25,0.75)--cycle;
    \draw[black,very thin,,fill=green!70,step=0.0625cm] (0.25,0.25) grid +(0.5,0.5)rectangle(0.25,0.25);
    \end{scope}
    
    \path [black,fill=cyan,thick,->,bend left]  (2,3.0) edge node[above] { \textit{RK stages} $B_l \subset B $=\{ $k\geq l_{max}$ to $l$\}}(5.0,3.0);
    \path [black,fill=cyan,thick,->,bend left] (5.0,1.0) edge node[below] { local \textit{blk\_sync}} (2,1.0);
    
    \end{scope}
    
    \begin{scope}[xshift=13cm,yshift=1.0cm,scale=0.025]
    \input{fig_tex/gr_im1_r1}
    \end{scope}
    \draw[thick,->] (12,1.75) -- (13,1.75) node[anchor=south east] {\texttt{zip}};
    \draw[thick] (6.5,0) --(6.5,5);
    \end{tikzpicture}
    }
    \caption{\small This figure illustrates the overview of time evolution using \uts~ and \nuts~ for multi-stage explicit timestepping, on octree grids. In order to perform FD computations, the adaptive grid is decomposed into uniform block patches with appropriate padding and spatial derivatives are evaluated on equispaced block representation(\textit{unzipped}) computed using previous timestep solution $u_{n}$. The \textit{unzip} operation results in a sequence of that which are used to compute the solution on the internal block (\textcolor{green!70}{$\blacksquare$}), using the padding values at the block boundary (\textcolor{yellow!30}{$\blacksquare$}). After time evolution, the next timestep $u_{n+1}$ is projected back to sparse grid  (\texttt{zip}) representation. In \uts, for each explicit stage, we evolve all the blocks using $\Delta t_{finest}$, followed by a global block synchronization operation. This global synchronization operation consists of projection of block local solution for \ziped~ representation (\cgn~), followed by inter-process communication and project back to unzipped representation. Note that \cgn~ are used for inter-process communication, since it is compact and does not contain node duplicate values. In \nuts, we have a block level loop which selects subset of blocks $B_l$ which are eligible to evolve, followed by the explicit stage loop. Once $B_l$ is evolved, we perform block synchronization for $B_l$, hence this is a local block synchronization. For this local block synchronization, for inter-process communication we use \dgn~ representation ({\it unzipped}) without padding region). Unlike \uts~ we cannot use \cgn~ representation, since blocks evolved are at different times. 
    \label{fig:overview}}
\end{figure}

\subsection{\uts: Global timestepping}
In this section, we describe how we perform global timestepping (\uts) on adaptive octrees with FD computations (see figure \ref{fig:overview}). In \uts, the finest space resolution will determine the timestep size for the entire grid, and the entire domain will march in a synchronized fashion. In most cases, this is not efficient especially when the overall percentage of finer octants across the grid is low, which is the case for many computational science applications \cite{einsteinathome,mantle,Fernando2018_GR}. Let us assume we have the solution $U^{n}$ defined on the adaptive grid. Then for each explicit stage, we unzip the intermediate timestep, loop over all the blocks to compute the block internal using FD stencils, perform zip operation and compute the next intermediate timestep(see algorithm \ref{alg:uts}). In order to synchronize the block padding regions we need inter-process communication, therefore before each unzip, we perform data exchange with neighboring processes using the compact \ziped~ representation. Note that all blocks do not depend on the values from neighboring processes. We label these as independent blocks, and those requiring values from other processes as dependent blocks. We exploit this property to overlap the \unzip~ computation with communication, i.e., we initiate the communication before starting the computation of padding for independent blocks, and upon receiving the data from neighboring processes perform the unzip operation for dependent blocks. 

\begin{algorithm}[tbh]
  \caption{Global timestepping (\uts)}\label{alg:uts}
  \footnotesize
  \begin{algorithmic}[1]
    \Require $U^{n}$ previous timestep, $\Delta t$, $B$ sequence of blocks
    \Ensure $U^{n+1}=U(t^n  \Delta t, \cdot)$
    \State $u\_unzip \leftarrow unzip(U^{n})$ 
    \For {$s=1:k$}
      \For {$b \in B$}
        \State $k[b,s] \leftarrow compute\_stage(u\_unzip, k[1,..,s-1])$ 
      \EndFor
    \State $k[s] \leftarrow blk\_sync(k[s])$
    \EndFor
    \State $U^{n+1} \leftarrow compute\_step(U^n,k)$
  \end{algorithmic}
\end{algorithm}

\subsection{\nuts: Local timestepping}
We now describe how to enable \nuts~ for explicit single and multi-stage timesteppers on 2:1 balanced  octree grids. 
Since $\Delta t = c \Delta x $, where $c$ is the CFL constant, we know that timestep sizes between coarser ($\Delta t_c$) and finer ($\Delta t_f$) grids are also 2:1 balanced, therefore, $\Delta t_c = 2\Delta t_f$. 

\subsubsection{Single stage explicit schemes} 
\begin{figure}[tbh]
 	\centering
 	\resizebox{0.8\columnwidth}{!}{
\begin{tikzpicture}[scale=1.0, every node/.style={scale=1.0}]
\tikzset{myptr/.style={decoration={markings,mark=at position 1 with %
    {\arrow[scale=2,>=stealth]{>}}},postaction={decorate}}}
    
    \draw[myptr] (0,0)--(10,0)  node[above] {$t_n$};

 	\draw[fill=blue!30] (0.5,0) circle [radius=0.1cm] node at (1.25,0.25) {$b_1$};
	\draw[fill=blue!30] (1.5,0) circle [radius=0.1cm] node at (3.0,0.25) {$b_2$};
	\draw[fill=blue!30] (3.5,0) circle [radius=0.1cm] node at (5.0,0.25) {$b_3$};
	\draw[fill=blue!30] (5.5,0) circle [radius=0.1cm] node at (7.2,0.25) {$b_4$};
	\draw[fill=blue!30] (7.5,0) circle [radius=0.1cm] ;

    \draw[black,myptr,dashed] (0,1)--(10,1)  node[above] {$t_{n+\frac{1}{4}}$};

    \draw[fill=blue!30] (0.5,2) circle [radius=0.1cm]; 
	\draw[fill=blue!30] (1.5,2) circle [radius=0.1cm];
	\draw[fill=blue!30] (3.5,2) circle [radius=0.1cm]; 
	\draw[fill=blue!30] (5.5,2) circle [radius=0.1cm];
	\draw[fill=blue!30] (7.5,2) circle [radius=0.1cm];

    \draw[myptr,thick,blue] (0.5,0) to[bend right,looseness=1.2] (0.5,1);  
    \draw[myptr,thick,blue] (1.5,0) to[bend right,looseness=1.2] (1.5,1);
    \draw[myptr,thick,blue] (3.5,0) to[bend right,looseness=1.2] (3.5,1);

    \draw[fill=blue!30] (0.5,4) circle [radius=0.1cm]; 
	\draw[fill=blue!30] (1.5,4) circle [radius=0.1cm];
	\draw[fill=blue!30] (3.5,4) circle [radius=0.1cm]; 
	\draw[fill=blue!30] (5.5,4) circle [radius=0.1cm];
	\draw[fill=blue!30] (7.5,4) circle [radius=0.1cm];

    \draw[myptr,thick,blue] (0.5,2) to[bend right,looseness=1.2] (0.5,3);  
    \draw[myptr,thick,blue] (1.5,2) to[bend right,looseness=1.2] (1.5,3);
    \draw[myptr,thick,blue] (3.5,2) to[bend right,looseness=1.2] (3.5,3);
    
    
    \draw[black,myptr] (0,2)--(10,2)  node[above] {$t_{n+\frac{1}{2}}$};


    \draw[myptr,thick,olive] (0.5,1) to[bend left,looseness=1.2] (0.5,2);  
    \draw[myptr,thick,olive] (1.5,1) to[bend left,looseness=1.2] (1.5,2);
    \draw[myptr,thick,olive] (3.5,0) to[bend left,looseness=1.2] (3.5,2);
    
    \draw[myptr,thick,blue] (5.5,0) to[bend right,looseness=1.2] (5.5,2);
    \draw[myptr,thick,blue] (7.5,0) to[bend right,looseness=1.2] (7.5,2);


    \draw[myptr,thick,olive] (0.5,3) to[bend left,looseness=1.2] (0.5,4);  
    \draw[myptr,thick,olive] (1.5,3) to[bend left,looseness=1.2] (1.5,4);
    \draw[myptr,thick,olive] (3.5,2) to[bend left,looseness=1.2] (3.5,4);
    \draw[myptr,thick,olive] (5.5,2) to[bend left,looseness=1.2] (5.5,4);
    \draw[myptr,thick,olive] (7.5,2) to[bend left,looseness=1.2] (7.5,4);

    
	
	\draw[black,myptr,dashed] (0,3)--(10,3)  node[above] {$t_{n+\frac{3}{4}}$};
	\draw[black,myptr] (0,4)--(10,4)  node[above] {$t_{n+1}$};
	
	\end{tikzpicture}}
 \caption{\small Simple illustration of local timestepping for single-stage schemes on 2:1 balanced adaptive grids. Here we have a refined block $b_1$ with neighbor $b_2$ that can be at most twice as large. We also consider block $b_3$ that doesn't have any neighbors smaller than itself. Note that 2:1 balancing ensures that these are the only cases that can exist for any adaptive mesh. Block $b_1$ and $b_3$ take timesteps corresponding to the size of the blocks. Block $b_2$ however takes $2\times$ as many timesteps compared to $b_3$, first a half-step to help its neighbor $b_1$ take its second timestep, and then a full size timestep to reach $t_{n+1}$. 
 \label{fig:nuts_single_stage}}	
\end{figure}
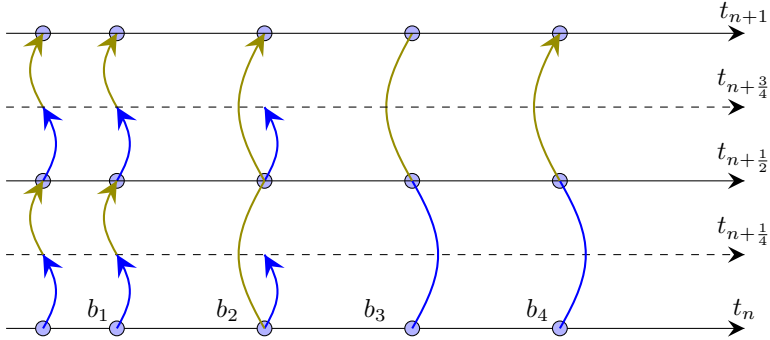

For single stage explicit schemes such as forward Euler, we can enable the correction between coarser ($b_c$) and finer ($b_f$) blocks, by making $b_c$ takes a pseudo timestep, for block $b_f$ (see figure \ref{fig:nuts_single_stage}). This approach is simple for single stage timesteppers, since only a single layer of coarser blocks that needs to take a pseudo timestep, as a correction for the finer blocks. Extending this to multi-step schemes is complicated and loses the local adaptive timestepping, since $N_s$ stage timestepper require $N_s$ layer of coarser blocks to take pseudo timesteps to make stage corrections. Therefore, we approach the multi-stage \nuts~ differently (see figure \ref{fig:overview}). 

\subsubsection{Multi-stage explicit schemes}

\begin{algorithm}[tbh]
  \caption{Local timestepping (\nuts)}\label{alg:nuts}
  \footnotesize
  \begin{algorithmic}[1]
    \Require $U^{n}$ previous timestep, $\Delta t$, $B$ sequence of blocks
    \Ensure $U^{n+1}=U(t^n  \Delta t, \cdot)$
    \State $V \leftarrow U^{n}$
    \State $u\_unzip \leftarrow unzip(U^{n})$ 
    \For {$l=l_{max}$ to $l_{min}$}
        \State $B_l\leftarrow compute\_blk\_subset(B,l)$
        \For {$b \in B_l$}
        \For {$s=1:k$}
            \State $k[b,s] \leftarrow compute\_stage(u\_unzip, k[1,..,s-1])$
            \State $k[s] \leftarrow blk\_sync\_local(k[s],B_l)$ 
        \EndFor
        \State $V \leftarrow compute\_step\_partial(V,k,B_l)$
    \EndFor
    \EndFor
    \State $U(t^{n}+\Delta t_{coarsest},\cdot) \leftarrow V$
  \end{algorithmic}
\end{algorithm}

In this section, we present how we perform \nuts~ for multi-stage explicit schemes (see algorithm \ref{alg:nuts}), on 2:1 balanced distributed octrees. The main challenge extending the single-stage timestepping approach for multi-stage schemes, is that the layers of blocks which have to perform pseudo timestep increases with the number of stages in the timestepping scheme. In order to avoid this, we need to decouple the stages $k_i$ from the timestep size used to evolve the solution. The linear relation (see equation \ref{eq:rk_crrection}) between stages $k_i$ and the $\partial^i _t U$ can be derived as described in \cite{liu2014nonuniform,liu2010nonuniform}, where $C$ is a lower triangular coefficient matrix, coefficients are derived from $a_{ij}$ coefficients of the explicit scheme. 

\begin{equation}
\label{eq:rk_crrection}
K = 
\begin{bmatrix}
k_{1}\\
k_{2}\\
\vdots \\
k_{p} 
\end{bmatrix} = P_{\Delta t}\times C \times \begin{bmatrix}
\partial_t U_{t=t^n}\\
\partial_t^2 U_{t=t^n}\\
\vdots \\
\partial_t^p U_{t=t^n} 
\end{bmatrix}
\end{equation}
where $P_{\Delta t}$ and $C$ are matrices defined as, 
\begin{equation}
\small
P_{\Delta t} =\begin{bmatrix}
1 & 0 & \cdots & 0 \\
0 & \Delta t & \cdots & 0 \\
\vdots  & \vdots  & \ddots & \vdots  \\
0 & 0 & \cdots & \Delta t^{p-1} 
\end{bmatrix}, C = \begin{bmatrix}
1 & 0 & \cdots & 0 \\
1 & c_{2,2} & \cdots & 0 \\
\vdots  & \vdots  & \ddots & \vdots  \\
1 & c_{p,2} & \cdots & c_{p,p} 
\end{bmatrix} 
\end{equation}

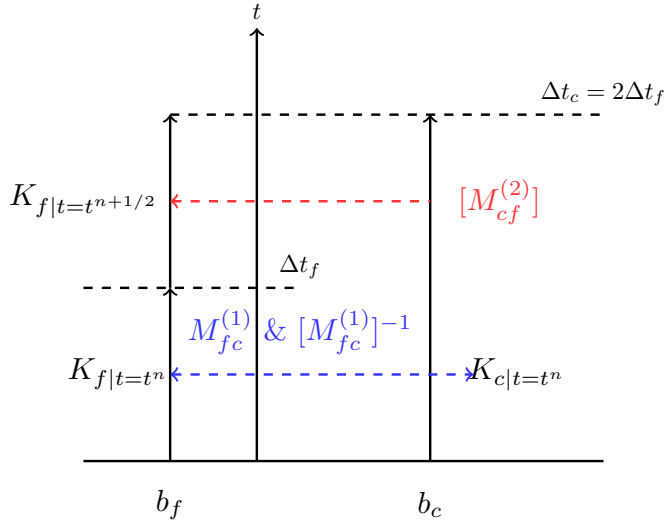
\begin{figure}[tbh]
	\centering
	\resizebox{0.7\textwidth}{!}{
	\begin{tikzpicture}
	\draw[black,thick,->] (2,0) -- (2,5)  node[above,scale=0.8] {$t$};
	\draw[black,thick,dashed] (1,4) -- (6,4) node[above,scale=0.8] {$\Delta t_c=2\Delta t_f$};
	\draw[black,thick,dashed] (0,2) -- (2.5,2) node[above,scale=0.8] {$\Delta t_f$};
	\draw[black,thick] (0,0) -- (2,0);
	\draw[black,thick] (2,0) -- (6,0);
	\node at (1,-0.5) {$b_f$};
	\node at (4,-0.5) {$b_c$};
	\draw[black,thick,->] (1,0) -- (1,2)  node at (0.4,1) {$K_{f|t=t^n}$};
	\draw[black,thick,->] (4,0) -- (4,4)  node at (5,1) {$K_{c|t=t^n}$};
	\draw[black,thick,->] (1,2) -- (1,4)  node at (0,3) {$K_{f|t=t^{n+1/2}}$};	
	\draw[thick,<->,blue!80,dashed] (1,1) -- (4.5,1)  node at (2.5,1.5) {$M^{(1)}_{fc}$ \& $[M^{(1)}_{fc}]^{-1}$};
	\draw[thick,<-,red!80,dashed] (1,3) -- (4.0,3)  node at(4.8,3) {$[M^{(2)}_{cf}]$}; 
	\end{tikzpicture}}
	\caption{\small Example of finer ($b_f$) to coarser ($b_c$) and coarser($b_c$) to finer ($b_f$) grid corrections. $M^{(1)}_{fc}$ is finer to coarser correction between fine step 1 and coarse step. $M^{(1)}_{fc}$ applied to finer grid we get the stage correction to coarser grids, and $[M^{(1)}_{fc}]^{-1}$ applied to coarser grid stages, we get the finer step 1 corrections. For the step 2 of the finer grid, we need the stage correction from coarser which is computed by applying, $[M^{(2)}_{cf}]$ to coarser step. }
	\label{fig:oct_2_1_nuts}
\end{figure}

Due to the 2:1 balancing, any given coarser and finer octant interfaces can differ at most by 1 refinement level. Hence timestep size ratio between coarser and finer levels are 2:1 balanced as well. Initially, we assume that all the blocks are synchronized in time. Let's consider adjacent  finer block $b_f$ and coarser block $b_c$ (see Figure~\ref{fig:oct_2_1_nuts}). Initially, both blocks have the solution at $t^n$. Both blocks compute the stage $k_i$ using their corresponding timestep size $\Delta t_f $ and $\Delta t _c$, but we need to apply a correction between the finer $K_f$ and coarser $K_c$ stages as per (\ref{eq:fc_1}), where $B_{dt}$ denotes an upper triangular matrix containing the coefficients computed from the Taylor expansion of each stage \cite{liu2014nonuniform}. 
\begin{equation}
\label{eq:fc_1}\small
K_{c|t=t^n} = P_{\Delta t_c} C B_{0} P^{-1}_{\Delta t_f} C^{-1} K_{f|t=t^n} = M^{(1)}_{fc} K_{f|t=t^n}
\end{equation}
Note that during the first step of the $b_f$ correction is applied to both coarse and finer blocks, after the first step of $b_f$, $b_c$ has already reached the coarser timestep, hence for the second step of $b_f$ the correction comes only from coarser grid (see equation (\ref{eq:fc_2}) and Figure \ref{fig:oct_2_1_nuts}).

\begin{equation}
\label{eq:fc_2}\small
K_{f|t=t^n + 1/2} = P_{\Delta t_f} C B_{\Delta_f} P^{-1}_{\Delta t_c} C^{-1} K_{c|t=t^n} = M^{(2)}_{cf} K_{c|t=t^n}
\end{equation}
More details on the error and stability analysis of these correction operators can be found in \cite{liu2014nonuniform,liu2010nonuniform}.

\subsubsection{Synchronization between blocks} 
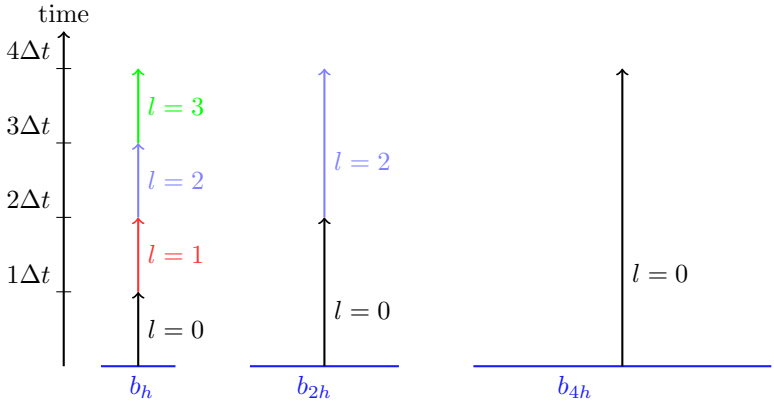
\begin{figure}[tbh]
    \centering
    \resizebox{0.8\columnwidth}{!}{
    \begin{tikzpicture}
        \draw [black,thick,->] (-0.5,0) -- (-0.5,4.5) node [above] {time};
        \newcommand*{\TickSize}{0.1cm}%
        \foreach \y in {1,...,4} {%
            \draw ($(-0.5,\y) + (-\TickSize,0)$) -- ($(-0.5,\y) + (\TickSize,0)$) node [near start, above left] {$\y \Delta t$};
        }
        \draw [thick,blue!90] (0,0) -- (1,0) node  [near start, below right] {$b_h$};
        \draw [thick,blue!90] (2,0) -- (4,0) node  [near start, below right] {$b_{2h}$};
        \draw [thick,blue!90] (5,0) -- (9,0) node  [near start, below right] {$b_{4h}$};
        
        \draw [black,thick,->] (0.5,0) -- (0.5,1) node [near start, above right] {$l=0$};
        \draw [black,thick,->] (3,0) -- (3,2) node [near start, above right] {$l=0$};
        \draw [black,thick,->] (7,0) -- (7,4) node [near start, above right] {$l=0$};
        
        \draw [thick,->,red!80] (0.5,1) -- (0.5,2) node [near start, above right] {$l=1$};
        \draw [thick,->,blue!50] (0.5,2) -- (0.5,3) node [near start, above right] {$l=2$};
        \draw [thick,->,blue!50] (3,2) -- (3,4) node [near start, above right] {$l=2$};
        
        \draw [thick,->,green] (0.5,3) -- (0.5,4) node [near start, above right] {$l=3$};
    \end{tikzpicture}}
    \caption{\small A simple illustration of block set $B=\{b_{h}, b_{2h}, b_{4h}\}$ that is evolving at each level $l$ of the time level loop. Assume in the beginning, all the blocks are synchronized in time. At level $l=0$, $B_0 =\{b_{h}, b_{2h}, b_{4h}\}$, perform timestep of its corresponding timestep size $\{\Delta t, 2\Delta t, 4\Delta t \}$. Note now the block $b_{4h}$ already reached the coarsest time over $B$. Similarly, for  $l=1$, $B_1=\{b_h\}$ , for $l=2$, $B_2=\{b_{h},b_{2h}\}$ and for $l=3$, $B_3=\{b_h\}$, are evolved with corresponding timestep sizes. After $B_3$, all the blocks have reached the next coarsest timestep size over $B$. Note that, appropriate time interpolation and corrections are applied across different refinement regions (see figure \ref{fig:oct_2_1_nuts}).}
    \label{fig:blk_loop}
\end{figure}

The computed correction operators allows us to relax the data dependency between adjacent blocks. The error caused by correction operators can be written as $\mathcal{O}(\Delta t) ^p$, where $p$ is the order of accuracy in the explicit timestepping method. Therefore it is important to make sure that these correction operators are computed between blocks where the time difference is minimal. For a given octree $T$ let $l_{max}$ and $l_{min}$ be the finest and the coarsest levels of refinement of the blocks. Then we evolve, the finest block with frequency of 1, next finest with frequency of $\frac{1}{2}$ and coarsest with the frequency of $\frac{1}{2^{\Delta L}}$ where $\Delta L = l_{max}- l_{min}$, until all the blocks reached the next coarsest time (see figure \ref{fig:blk_loop}). The above with 2:1 balancing the maximum time gap between two adjacent blocks ($b_f$, $b_f$) is bounded by $\Delta t_f$, the step-size of the finer block. Once all the blocks that can be processed have been computed and stored in the \dgn~ representation (see figure \ref{fig:dg_to_cg}), we perform data exchange between partitions to update ghosted values. Note that unlike \uts, we cannot use the \cgn~ representation for communication  since different grid points can be at a different time during \nuts. Therefore we need duplicate degrees of freedoms at octant level to perform \nuts~ ghost synchronization. Following ghost synchronization, we perform the block-wise correction as per (\ref{eq:fc_1}) and (\ref{eq:fc_2}). These are applied and copied to the block padding region (see figure \ref{fig:oct_2_1_nuts}) of the neighboring blocks. 

\subsubsubsection{Partial block synchronization} At a given time level $l$, we evolve a subset of blocks $B_l$, hence at the time of the inter-process synchronization (ghost exchange), we only need to communicate data corresponding to $B_l$. We pre-compute maps for each time level to perform partial ghost synchronization, to reduce the amount of data communicated between processors. This is more efficient than the global block synchronization performed in \uts, and in addition to the reduction of work by reducing the number of timesteps taken by coarser regions, it also reduces the amount of data synchronization needed. 

{}

\subsubsection{Weighted partitioning} 
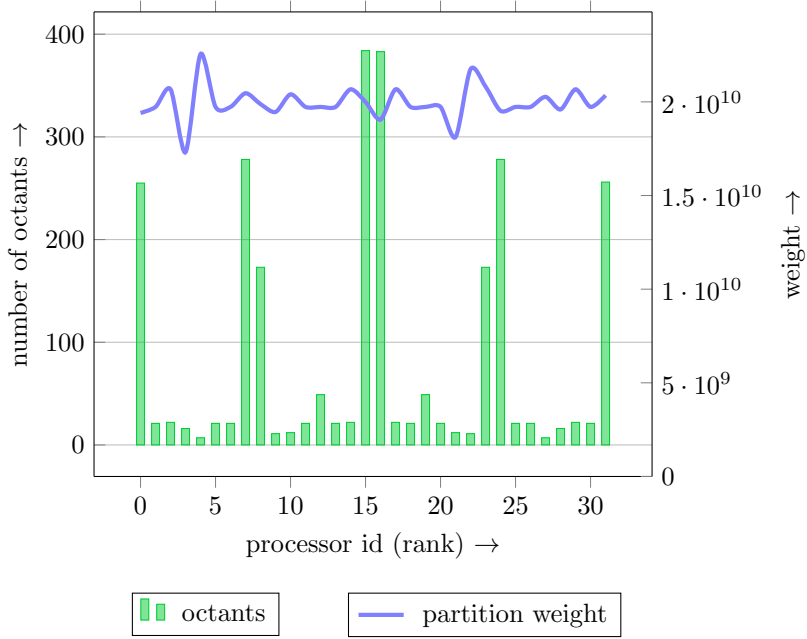
\begin{figure}[tbh]
	\centering
\begin{tikzpicture}
    \begin{axis}[
        width  = 9cm,
        axis y line*=left,
        ybar=5*\pgflinewidth,
        bar width=3pt,
        ymajorgrids = true,
        ylabel = {number of octants   $\rightarrow$},
        xlabel = {processor id (rank) $\rightarrow$},
        legend style={
                at={(0.2,-0.25)},
                anchor=north,
                column sep=1ex
        }
    ]
    \addplot[style={green!60!teal,fill=green!60!teal,fill opacity=0.5,mark=none}]
           table[x={rank}, y = {num_ele}]{dat/w_part_chpc.dat};
    \legend{octants}
    \end{axis}
    \begin{axis}[
    axis y line*=right,
    axis x line=none,
    width = 9cm,
    ylabel = {weight $\rightarrow$},
    scaled y ticks = false,
    ymin=0,
    legend style={
        at={(0.7,-0.25)},
        anchor=north,
        column sep=1ex
    }
    ]
    \addplot[blue!50,ultra thick, smooth] table[x={rank}, y ={weight}]{dat/w_part_chpc.dat};
    \legend{partition weight}
    \end{axis}
\end{tikzpicture}
\caption{A simple illustration of our weighted SFC-based partitioning scheme for an adaptive octree partitioned across 32 processors. The bar plot shows the number of octants each partition has, and the line plot represents the {\bf total} weight of each partition. Note that partitions with low octant counts have highly refined regions; hence they need to perform a larger number of timesteps to reach the coarser level timestep and vice versa, but the total weight of each partition is roughly equal, i.e., all partitions perform roughly the same amount of work.}
\label{fig:w_part}
\end{figure}

We use SFC based partitioning scheme to distribute work among the processors. For each local partition $\tau_k$, the amount of work that each block has to perform to reach to the global coarsest timestep depends on the block refinement level. For example a block of level $l$ has to perform \texttt{2x} timesteps compared to its adjacent coarser block at level $l-1$. A partitioning scheme with equal weights assigned to each octant will result in load imbalanced partitions for \nuts~ timestepping. In order to overcome that, we perform weighted partitioning, where the relative weight of block increases with the refinement level (see Figure \ref{fig:w_part}). We modified our SFC-based partitioning scheme to account for the specified weights of the blocks. A 3d SFC curve can be considered to be an injective mapping between 1d domain to 3d octree domain. Using SFC ordering, we can sort the octants which results in a linear ordering of the octants, while ensuring spatial locality.
Once ordered according to the SFC, partitioning the domain reduces to partitioning a $1D$ curve. In weighted SFC-partitioning, we use weighted length of the curve i.e. we aim for equal $\sum_{e\in \tau_k} w_e$, where $w_e$ denotes the weight of the octant, $\tau_k$ a given partition. Note that for \uts, we use $w_e = 1, \forall e \in \tau_k$.


\subsection{\uts~ Vs. \nuts: Approximating the speedup} \label{subsec:appx_speedup}
Here we present theoretical bounds for the work performed by the \uts~ and \nuts~ timesteping approaches, on 2:1 balanced octrees. Let $L=(l_{max}-l_{min})$ be the difference between the maximum and minimum refinement levels for a given octree.  Let $W = \{\alpha_0,...,\alpha_L\}$, be the corresponding work for a block sequence $B=\{b_l\}_{l=l_{min}}^{l_{max}}$, then for matching the finest timestep sizes between \uts~ and \nuts~ finest levels, the work performed by the \uts~ and \nuts~ schemes can be written as (\ref{eq:w_uts}) and (\ref{eq:w_nuts}). Assuming constant time to perform a work unit, for speedup $S$, $1/S$ can be written as (\ref{eq:invS}). Since, $\frac{\alpha_0}{|W|} < \frac{1}{S}$, maximum theoretical speedup that can achieved, for a given block distribution can be written as $ S < \frac{|W|}{\alpha_0}$.

\begin{align}
    W_{lts} &= \sum_{l=0}^{L} 2^{L-l}\alpha_l \label{eq:w_nuts}\\
    W_{gts}  &= 2^L \sum_{l=0}^{L} \alpha_l \label{eq:w_uts} \\
    \frac{W_{lts}}{W_{gts}} & = \frac{1}{|W|} (\alpha_0 + \frac{\alpha_1}{2} + ... +\frac{\alpha_L}{2^{L}} ) < 1 \label{eq:invS} \text{ Where, } \\
    |W| &= \sum_{l=0}^{L} \alpha_l 
\end{align}

\section{Results}
\label{sec:results}
\subsection{Experimental setup}

The large scalability experiments reported in this paper were
performed on TACC's \Frontera~ supercomputer. \Frontera~ is an Intel supercomputer at Texas advanced computing center (TACC) with a total of 8,008 nodes, each consisting of a Xeon Platinum 8280 ("Cascade Lake") processor, with a total of 448,448 cores. Each node has 192GB of memory. The interconnect is based on Mellanox HDR technology with full HDR (200 Gb/s) connectivity between the switches and HDR100 (100 Gb/s) connectivity to the compute nodes.
%
%

\subsection{Non-linear and linear wave propagation}
In this section, we introduce a simple model to demonstrate \nuts, the classical wave equation. We write the classical wave equation in a form with first derivatives in time and second derivatives in space. This allows us to easily apply the specified, explicit schemes. 

For a scalar function $\chi(t,x^i)$, the classical wave equation in Cartesian coordinates $(t,x,y,z)$ with a non-linear source term can be written as, 
\begin{equation}
\label{eq:lsm}
\frac{\partial^2\chi}{\partial t^2} 
 - \left(\frac{\partial^2}{\partial x^2} + \frac{\partial^2}{\partial y^2}
     + \frac{\partial^2}{\partial z^2}\right) \chi  = 
-c \frac{\sin(2\chi)}{r^2},
\end{equation}
where $r = \sqrt{x^2 + y^2 + z^2}$. We write the equation as first order
in time system by introducing the variable $\phi$ as
\begin{align}
	\frac{\partial\chi}{\partial t} &= \phi \label{eq:chi} \\
	\frac{\partial\phi}{\partial t} &= \left(\frac{\partial^2}{\partial x^2} 
         + \frac{\partial^2}{\partial y^2} 
         + \frac{\partial^2}{\partial z^2}\right) \chi 
       -c \frac{\sin(2\chi)}{r^2} 
       \label{eq:phi}.
\end{align}
For non-linear wave propagation results presented in this paper, we used $c=1$.
We choose outgoing radiative boundary conditions for this system~\cite{Alcubierre:1138167}. We assume that the variables $\chi$ and
$\phi$ approach the form of spherical waves as $r\to\infty$, which decay as $1/r^k$. The radiative boundary conditions then have the form
\begin{equation}
	\frac{\partial f}{\partial t} = \frac{1}{r}
\left(     x\frac{\partial f}{\partial x} 
         + y\frac{\partial f}{\partial y} 
         + z\frac{\partial f}{\partial z}\right) - k(f - f_0),
\end{equation}
where $f$ represents the functions $\chi$ and $\phi$, and $f_0$ is
an asymptotic value. We assume $k=1$ for $\chi$ and $k=2$ for $\phi$.

For the linear wave propagation results presented, we simply zero out the non-linear source term (i.e. set $c=0$). The analytical solution for the 1d linear wave operator in (\ref{eq:lsm}) with zero source term can be written as, 
\begin{equation}
    \chi(t,x) = \frac{f(x-t) + f(x + t)}{2}, \text{ where } \chi(0,x) =f(x)
\end{equation}

\begin{figure}[tbh]
	\includegraphics[width=0.32\columnwidth]{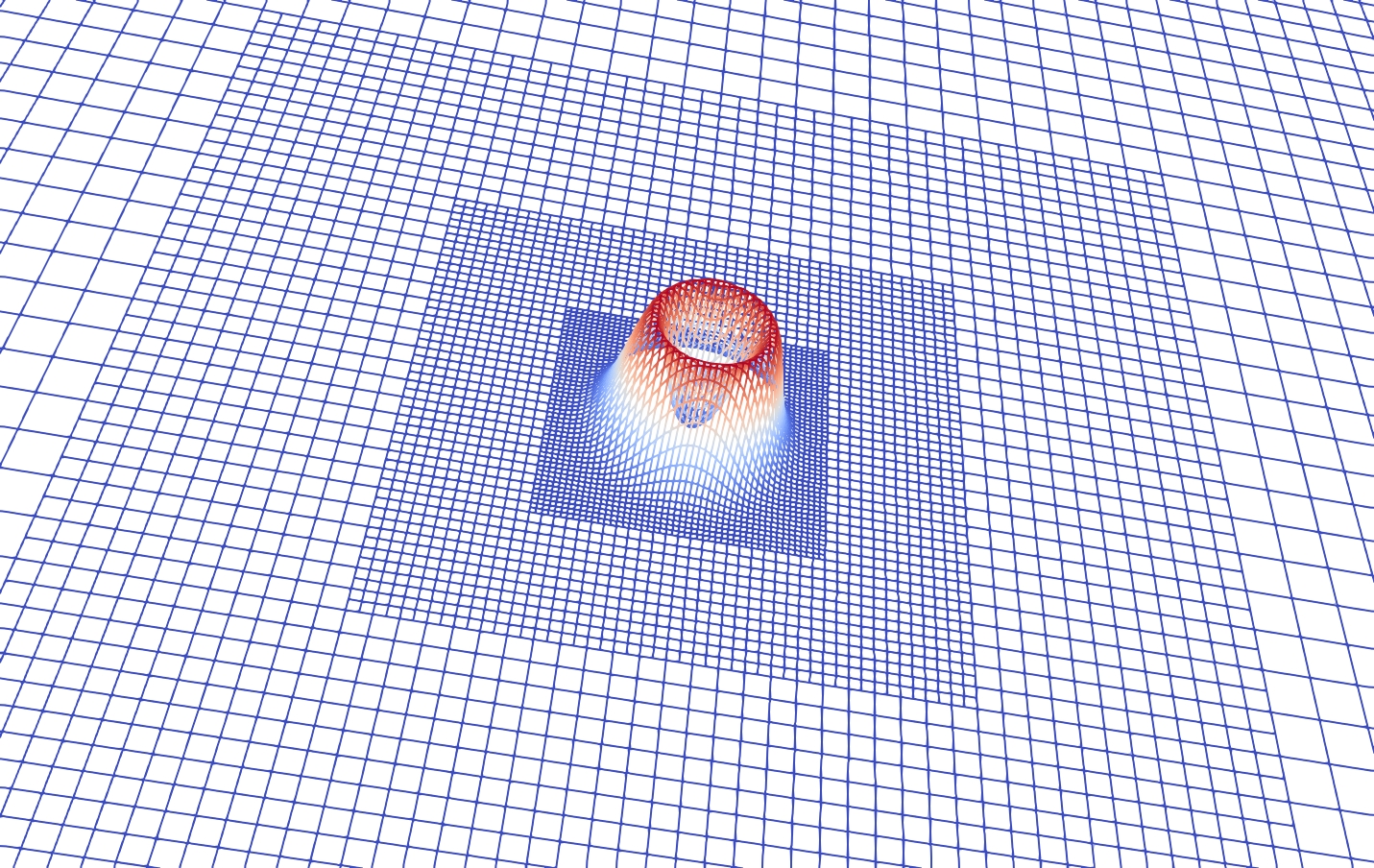}
	\includegraphics[width=0.32\columnwidth]{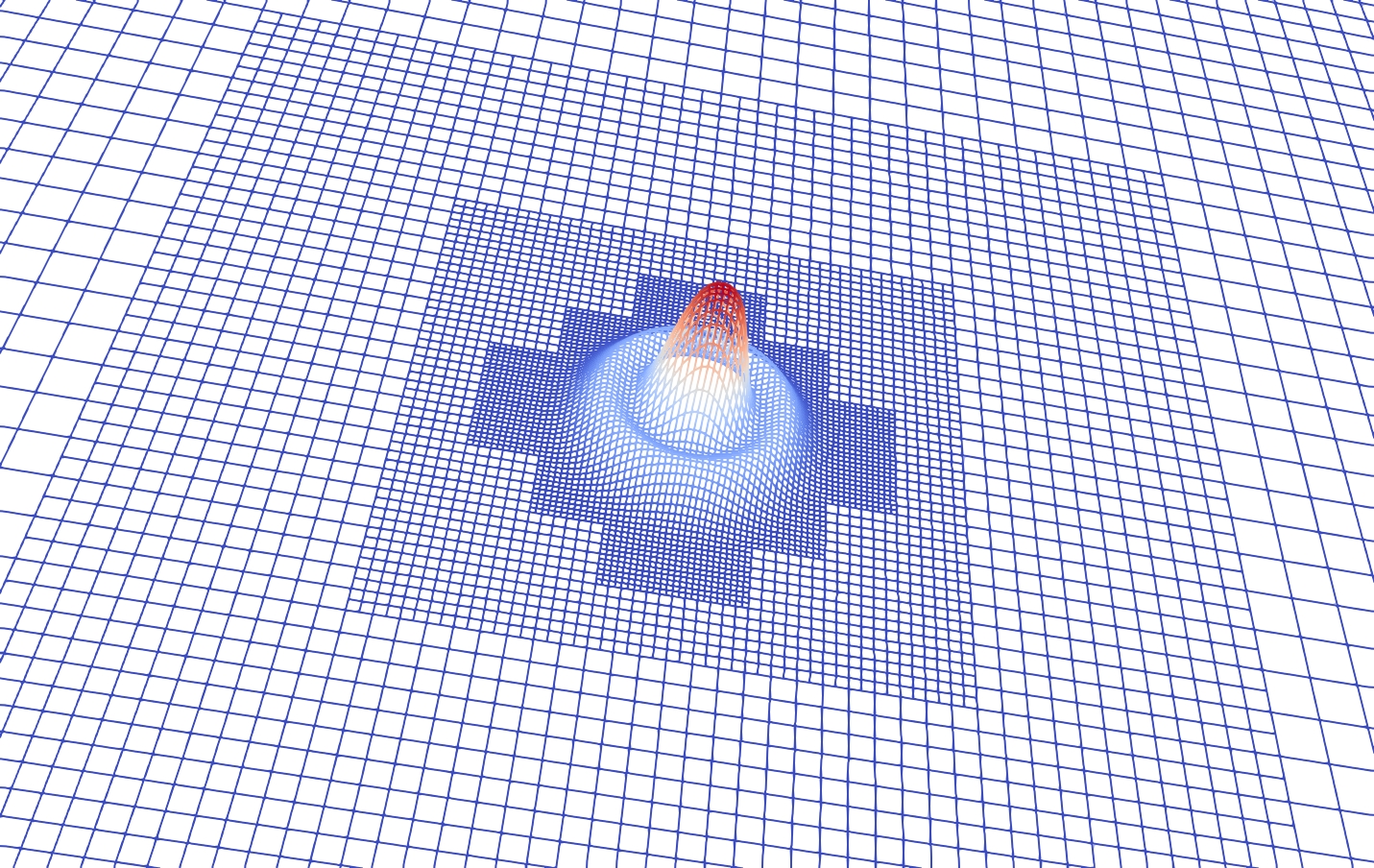}
	\includegraphics[width=0.32\columnwidth]{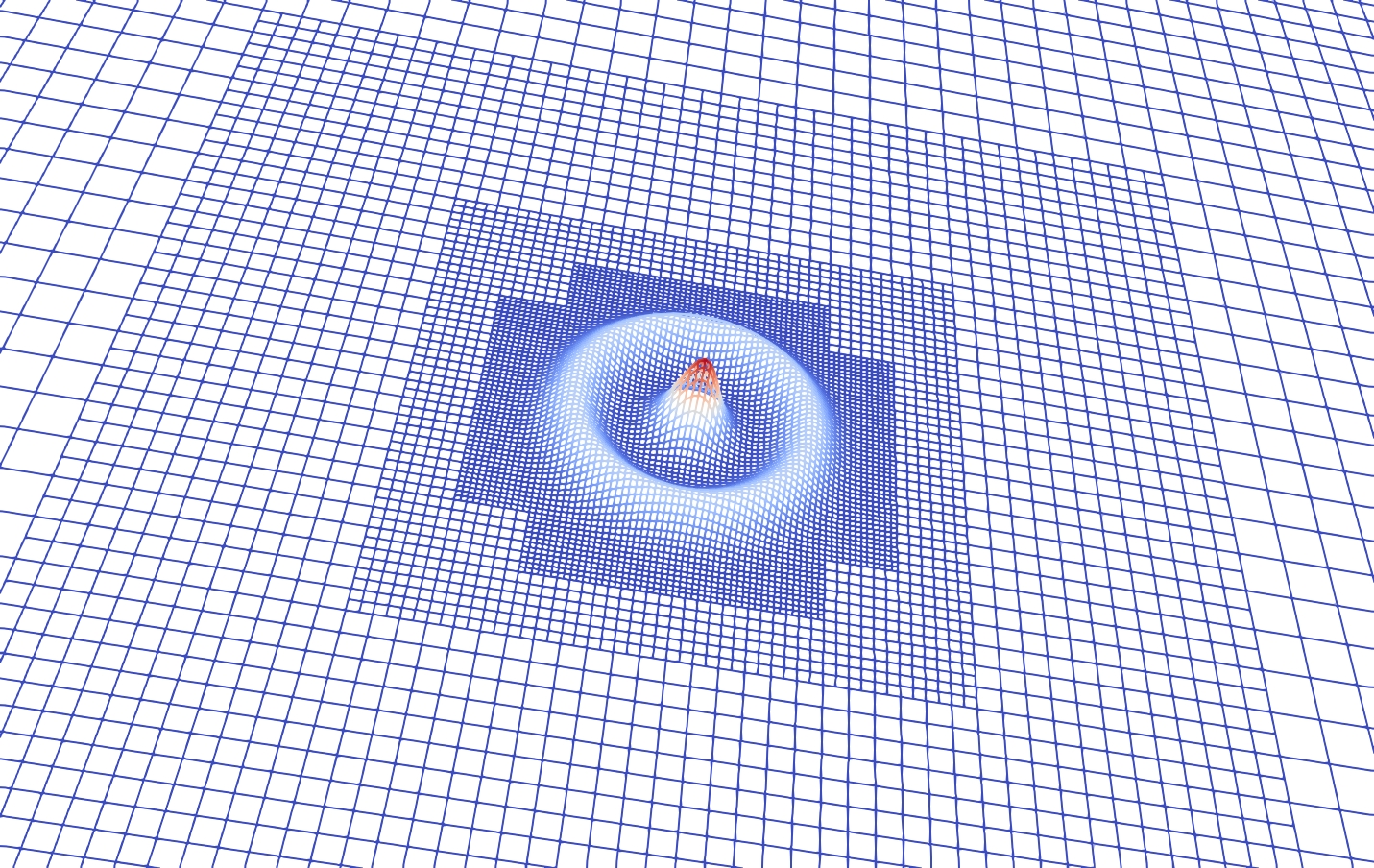}
	
	\includegraphics[width=0.32\columnwidth]{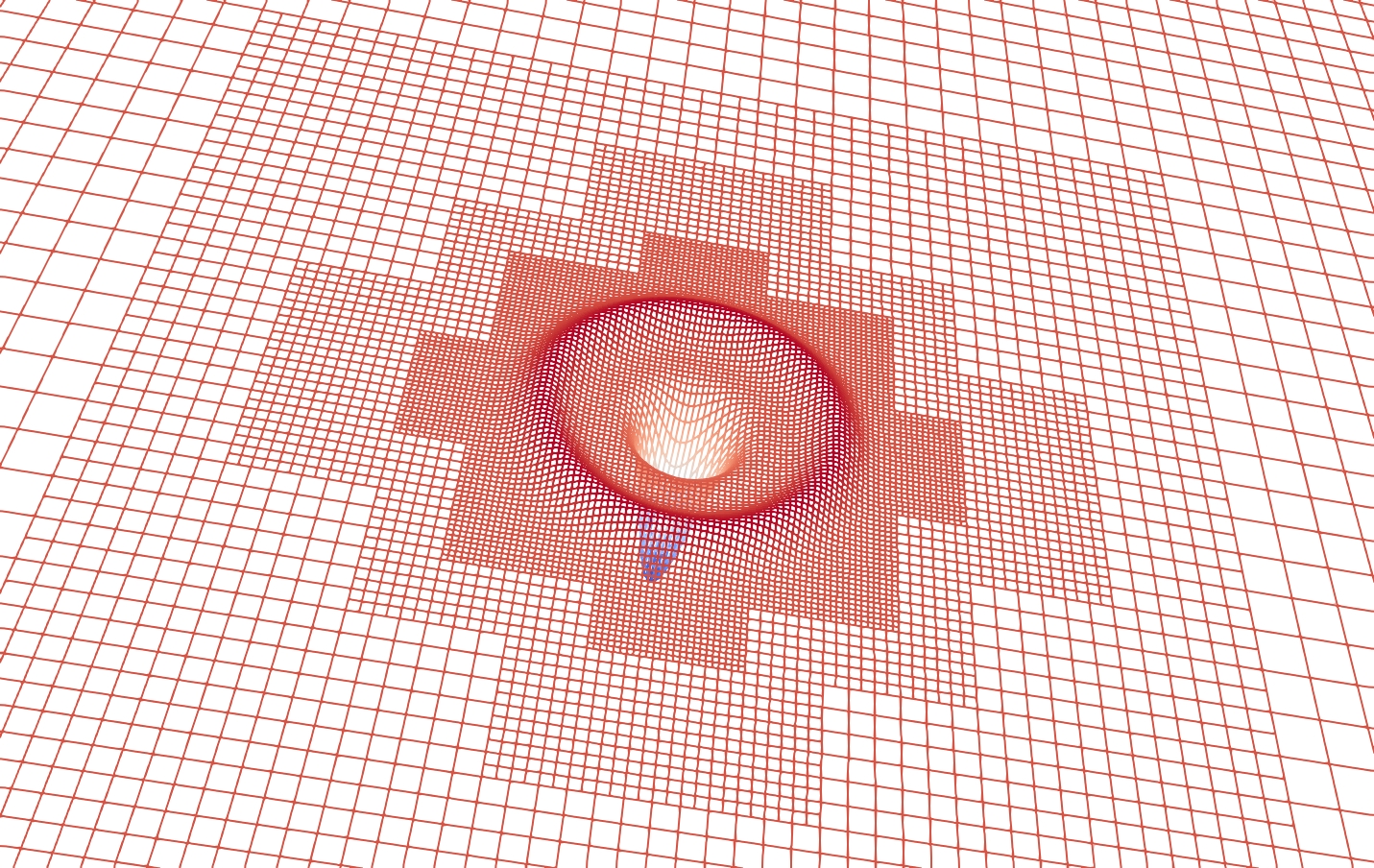}
	\includegraphics[width=0.32\columnwidth]{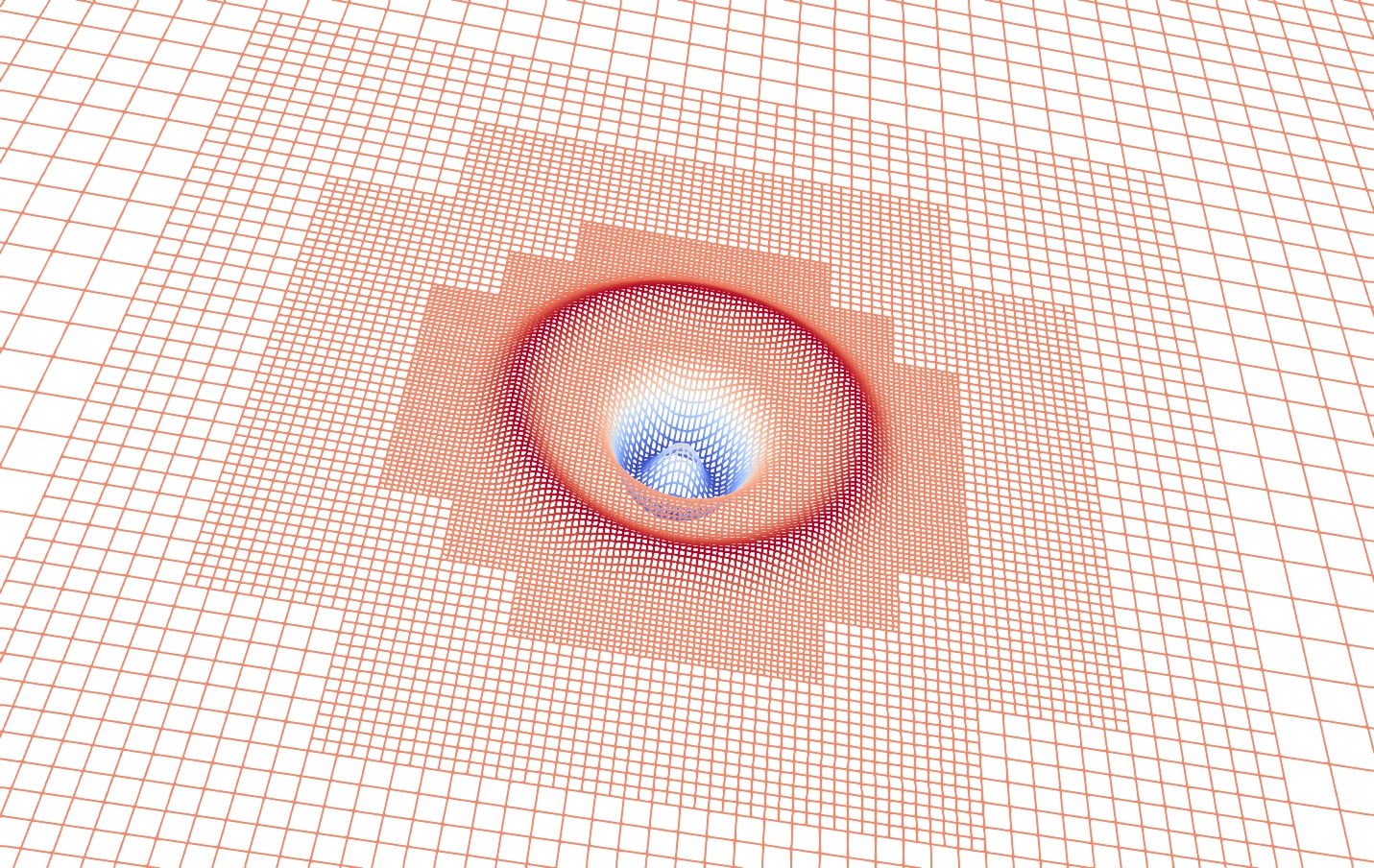}
	\includegraphics[width=0.32\columnwidth]{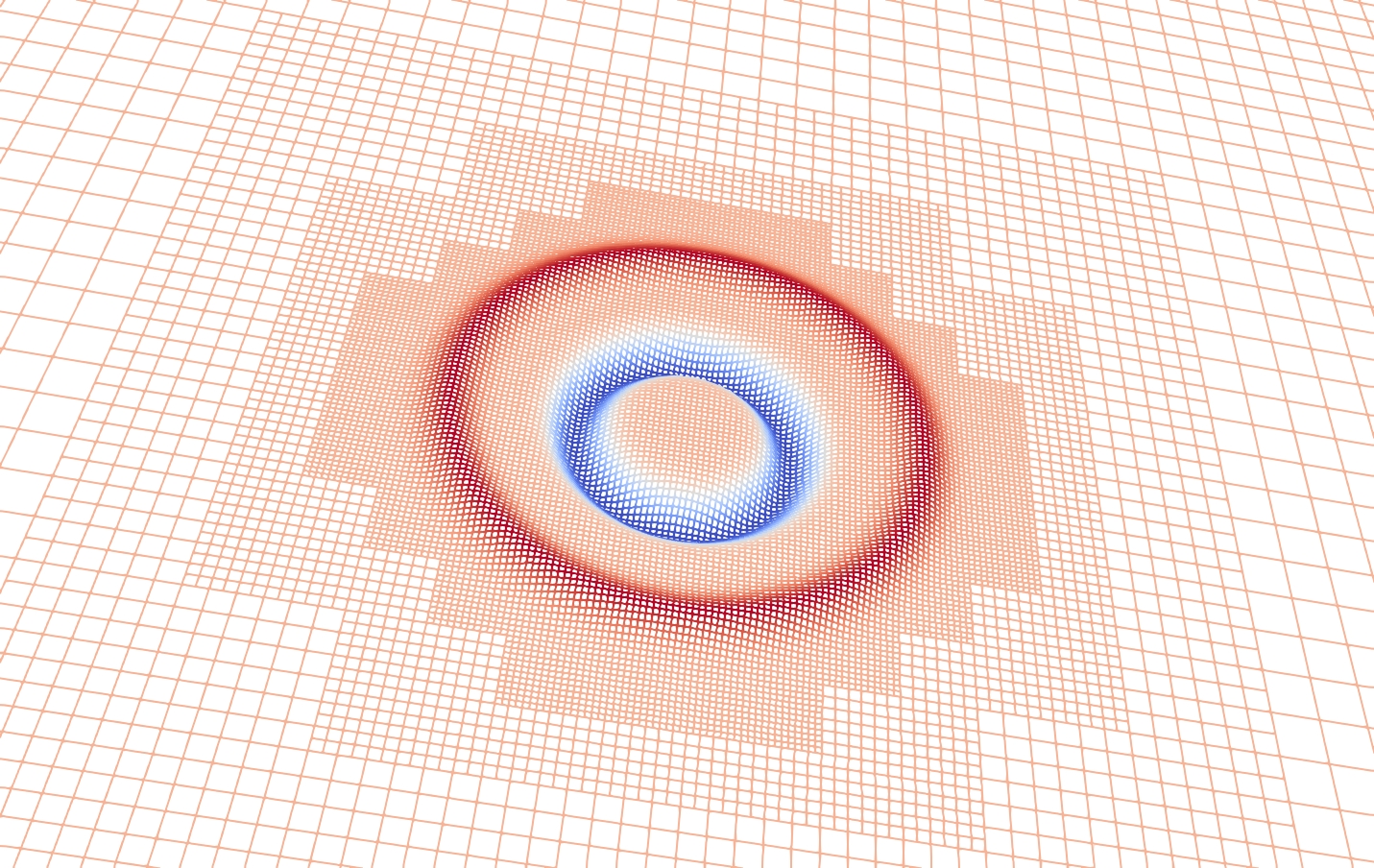}
	\caption{\small Plots for linear wave propagation with \nuts\ with a velocity vector $(1,1,1)$ using a Gaussian pulse centered at $(0,0,0)$ as the initial condition. Images shown from left to right and top to bottom in increasing order of simulation time. 
	\label{fig:lsm}}
\end{figure}

\subsection{Accuracy}

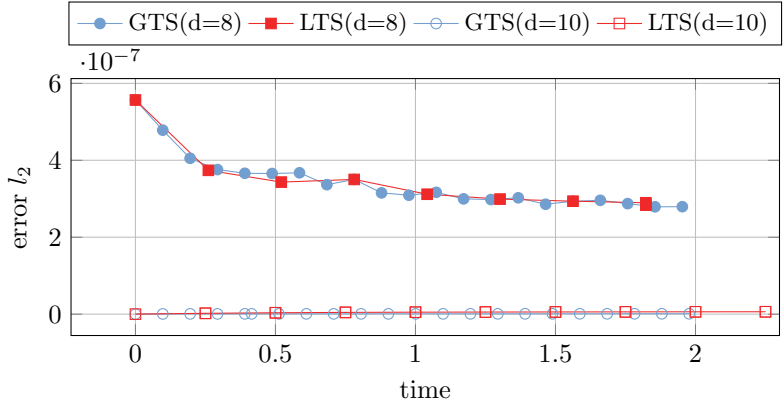
\begin{figure}[tbh]
    \centering
	\begin{tikzpicture}
	\begin{axis}[height=5cm, width=11cm, grid=major, xlabel={time}, ylabel={error $l_2$}, legend style={at={(0.5,1.3)}, anchor=north},legend columns=4,grid=major, xmax=2.3]
	\addplot[color=skyblue1,mark=*] table[x={time},y={l2}] {dat/nlsm_ets_d8_error_red.dat};
	\addplot[color=scarletred1,mark=square*] table[x={time},y={l2}] {dat/nlsm_nuts_d8_error_red.dat};
	\addplot[color=skyblue1,mark=o,mark options=solid] table[x={time},y={l2}] {dat/nlsm_ets_d10_error_red.dat};
	\addplot[color=scarletred1,mark=square,mark options=solid] table[x={time},y={l2}] {dat/nlsm_nuts_d10_error_red.dat};
	\legend{\small GTS(d=8), \small LTS(d=8), \small GTS(d=10), \small LTS(d=10)}
	\end{axis}
	\end{tikzpicture}
	\caption{Discrete $l_2$ error compared to the analytical solution for the 1d wave operator in 3d for \gts~ and \nuts~ timestepping. For this experiment we used \maxDepth~ of 8 and 10 with refinement trigger tolerance of $10^{-5}$.  \label{fig:accu_d8}}
\end{figure}

We conduct numerical experiments using linear and non-linear wave propagation to test the accuracy of our methods and implementation. For linear wave propagation, we compute the analytical solution for wave propagation in the $x$ direction and compare the analytical solution with the computed solution using global and local timestepping. For the above experiments, we used a third-order RK scheme with increasing spacetime resolution. Figure \ref{fig:accu_d8} shows numerical error for \lts~ and \gts~ approaches with increase resolution. 

Since the computation of the analytical solution for the 1d wave operator with non-linear source term is complicated, we compare the $l_\infty$ norm computed on the numerical difference between global and local evolved timesteps. In table \ref{tab:accu_nl} we present the difference between the solution $\chi$ evolved using \gts~ and \lts~ for increasing maximum refinement level(\maxDepth) 8 and 10. Again, both \uts\ and \nuts\ are in agreement to machine precision. This demonstrates that we can use \nuts\ in lieu of \uts\ without sacrificing accuracy or stability for both linear as well as non-linear problems. 

\begin{table}[tbh]
  \centering
  \renewcommand{\arraystretch}{1.2}
  \begin{tabular}{|p{2cm}|p{3cm}|p{3cm}|}
    \hline
    \multirow{1}{2cm}{Time(s)} & \multicolumn{2}{c|}{$||\chi_{lts}-\chi_{gts}||_\infty$} \\
    \cline{2-3}
     & \maxDepth=8 & \maxDepth=10 \\
    \hline
0.000000    &   0   &	0   \\
0.130208    &   9.93E-39	&   2.22E-16 \\
0.260417    &   6.92E-34    &   1.11E-15 \\
0.390625    &   4.86E-30	&   1.50E-15 \\
0.520833    &   6.20E-27	&   2.08E-15 \\
0.651042    &   1.39E-17	&   5.11E-15 \\
0.781250    &   5.55E-17	&   1.09E-14 \\
0.911458    &   1.39E-16	&   9.69E-15 \\
1.041670    &   2.09E-16	&   9.21E-15 \\
\hline
\end{tabular}
  \caption{$l_\infty$ difference between \uts~ and \nuts~ at corresponding timesteps for non-linear wave propagation. For this experiment we used \maxDepth~ of 8 and 10 with refinement trigger tolerance of $10^{-5}$. Note that for \maxDepth~ 8, 8 global timesteps and for \maxDepth~ 10, 32 global timesteps were equivalent for a single \lts~ step. \label{tab:accu_nl}}
\end{table}


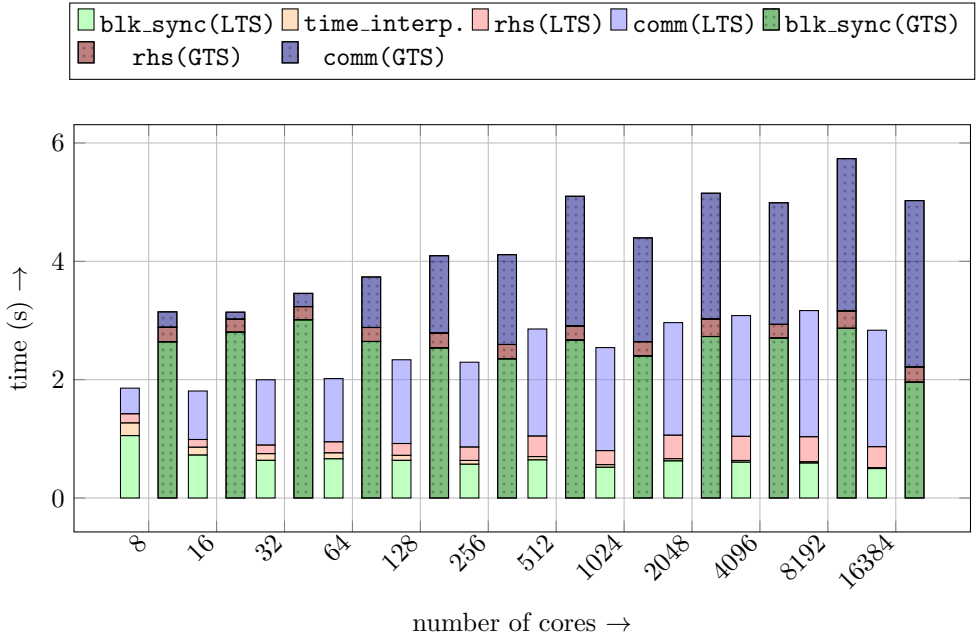
\begin{figure*}[tbh]
	\centering
	\begin{tikzpicture}[draw=black]
	\begin{axis}[
	ybar stacked, bar width=0.25cm,
	xlabel={number of cores $\rightarrow$},
	ylabel={time (s) $\rightarrow$ },symbolic x coords={8,16,32,64,128,256,512,1024,2048,4096,8192,16384},width=13.5cm,height=7cm,
	xtick = data, 
	x tick label style={rotate=45, anchor=east, align=right},
	legend style={at={(0.5,1.3)}, anchor=north},legend columns=5,grid=major]
	\addplot [fill=green!50,  fill opacity=0.5] [bar shift=-.25cm] table[x={act_npes}, y = {blk_unzip}]{dat/enuts_ws1_frontera.dat};
	\addplot [fill=orange!50, fill opacity=0.5] [bar shift=-.25cm] table[x={act_npes}, y = {nuts_correct}]{dat/enuts_ws1_frontera.dat};
	\addplot [fill=red!50, fill opacity=0.5] [bar shift=-.25cm] table[x={act_npes}, y = {rhs}]{dat/enuts_ws1_frontera.dat};
	\addplot [fill=blue!50, fill opacity=0.5] [bar shift=-.25cm] table[x={act_npes}, y = {comm}]{dat/enuts_ws1_frontera.dat};
	\resetstackedplotsOne
	\addplot [fill=green!50!black, fill opacity=0.5, postaction={pattern=dots}] [bar shift=.25cm] table[x={act_npes}, y = {unzip32}]{dat/ets_ws1_frontera.dat};
	\addplot [fill=red!50!black  , fill opacity=0.5, postaction={pattern=dots}] [bar shift=.25cm] table[x={act_npes}, y = {rhs32}]{dat/ets_ws1_frontera.dat};
	\addplot [fill=blue!50!black , fill opacity=0.5, postaction={pattern=dots}] [bar shift=.25cm] table[x={act_npes}, y = {comm}]{dat/ets_ws1_frontera.dat};
	\legend{\texttt{blk\_sync(LTS)},\texttt{time\_interp.}, \texttt{rhs(LTS)}, \texttt{comm(LTS)}, \texttt{blk\_sync(GTS)}, \texttt{rhs(GTS)}, \texttt{comm(GTS)}  }
	\end{axis}
	\end{tikzpicture}
	\caption{\small Weak scaling results on TACC's \Frontera ~ for RK3 timestepping using \lts (left) and \gts (right) approaches. For this experiment, the maximum and minimum refinement levels are $l_{max}=8$ and $l_{min}=3$, hence $\Delta L=5$. Therefore a single \nuts~ step is equivalent to $2^{\Delta L}=32$ global timesteps. For \lts, the plot shows the overall cost breakdown between block synchronization (\texttt{blk\_sync}), applying time interpolations between blocks (\texttt{LTS\_interp.}), computing the right hand side (\texttt{rhs}) and communication costs (\texttt{comm}). For \gts, we show the cost breakdown between \texttt{blk\_sync}, \texttt{rhs} and \texttt{comm}. Note that for \gts~ time interpolations are not required. Note the significant difference of \texttt{blk\_sync} cost between \lts~ and \gts. For \gts~ \texttt{blk\_sync} is a global operation, while in \lts~  \texttt{blk\_sync} is a local operation, where synchronization performed only on the subset of blocks, which are currently being evolved. These weak scaling results were performed using a grain size of $\sim 100K$ unknowns per core, with the number of cores ranges from $8$ to $16,384$ cores. 
	The largest problem recorded had $1.6\times 10^9$ unknowns. The above results are generated for radial wave propagation with  a \maxDepth~ 10 and a refinement tolerance of $10^{-5}$.}\label{fig:ws_enuts_g300} 
\end{figure*}

\subsection{\nuts~ efficiency and space adaptivity}
As mentioned in \S\ref{subsec:appx_speedup}, we can approximate the speed up $S$ between \nuts~ and \uts~ for a given octant distribution. 
Since we can end up with meshes where the use of \nuts\ will not provide significant advantages over \uts, we can selectively use \nuts\ based on the expected speed up. To evaluate our speed up model (\ref{eq:invS}), and to assess the overhead of applying the \nuts\ correction operators, 
we computed the actual speed up reported for the linear wave propagation with increasing \maxDepth. The estimated and reported speed up values are presented in the table \ref{tab:s_lsm}. As can be seen, the estimated speed up values are sufficiently close to the predicted ones, allowing applications to determine when it is beneficial to use \nuts.

\begin{table}[tbh]
    \centering
    \resizebox{\columnwidth}{!}{
    \begin{tabular}{ |c|c|c|c|c|c|c|} 
        \hline
        d & $l_{min}$ & $l_{max}$ & \uts(s)& \nuts(s) & est. speed up & reported speed up\\
        \hline
        9  &	 2  &	7  &          2.91   &	0.88 &	3.31   & 3.30 \\
        10 &     3 	&   8  &	     26.69	  &   9.26  &  3.31   &	2.87 \\
        11 &	 2	&   9  &	    114.74   &	  36.76  &	3.49   &	3.12 \\
        12 &	 2	&   10 &	    238.05 &	 73.94  &	3.56 &	3.21 \\
        \hline
    \end{tabular}}
\caption{The estimated vs. reported speed up for \lts~ over \gts~ for linear wave propagation with increasing \maxDepth~ for adaptive octrees.
}
\label{tab:s_lsm}
\end{table}{}

\begin{figure}[tbh]
	\includegraphics[width=0.32\columnwidth]{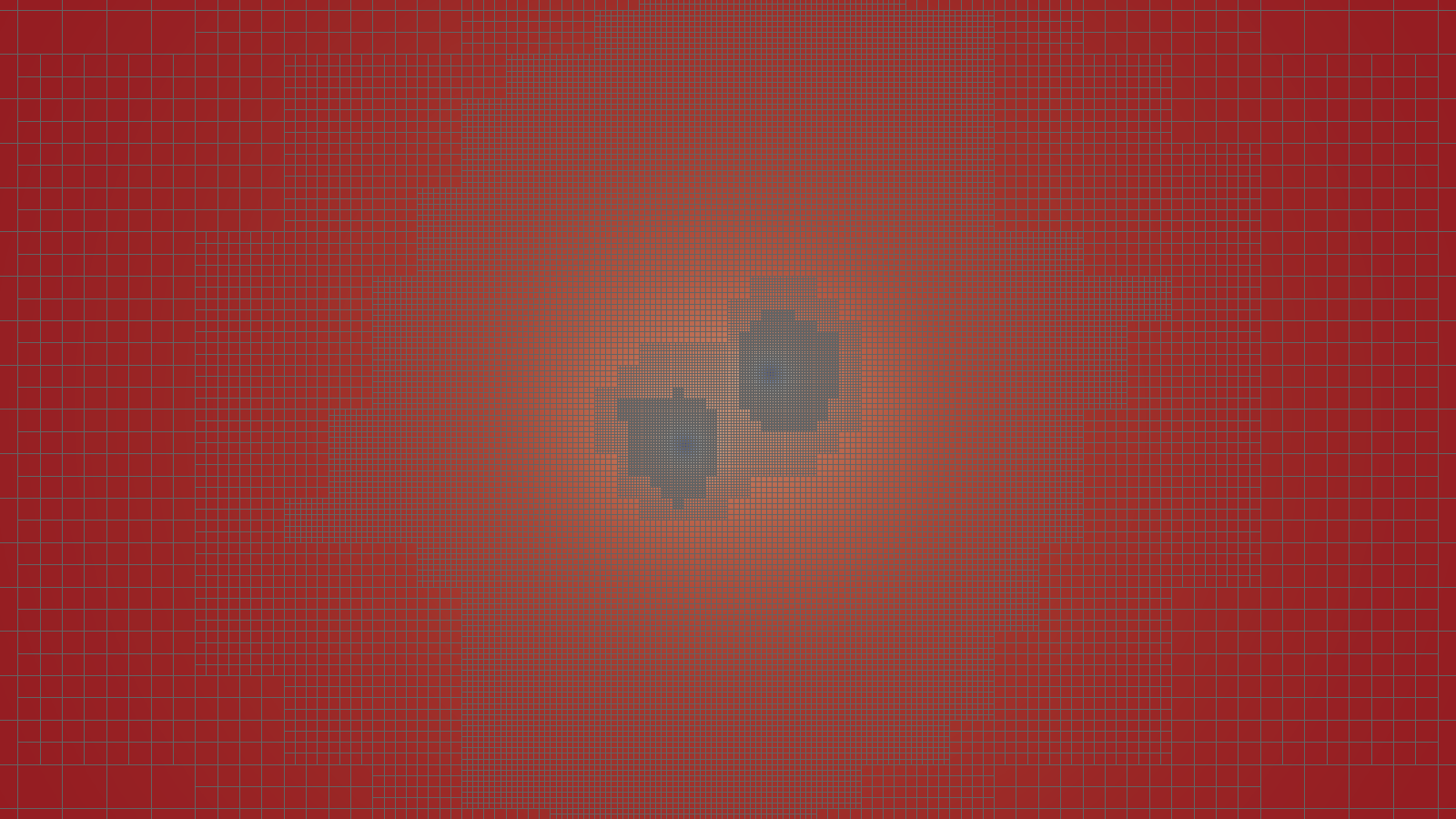}
	\includegraphics[width=0.32\columnwidth]{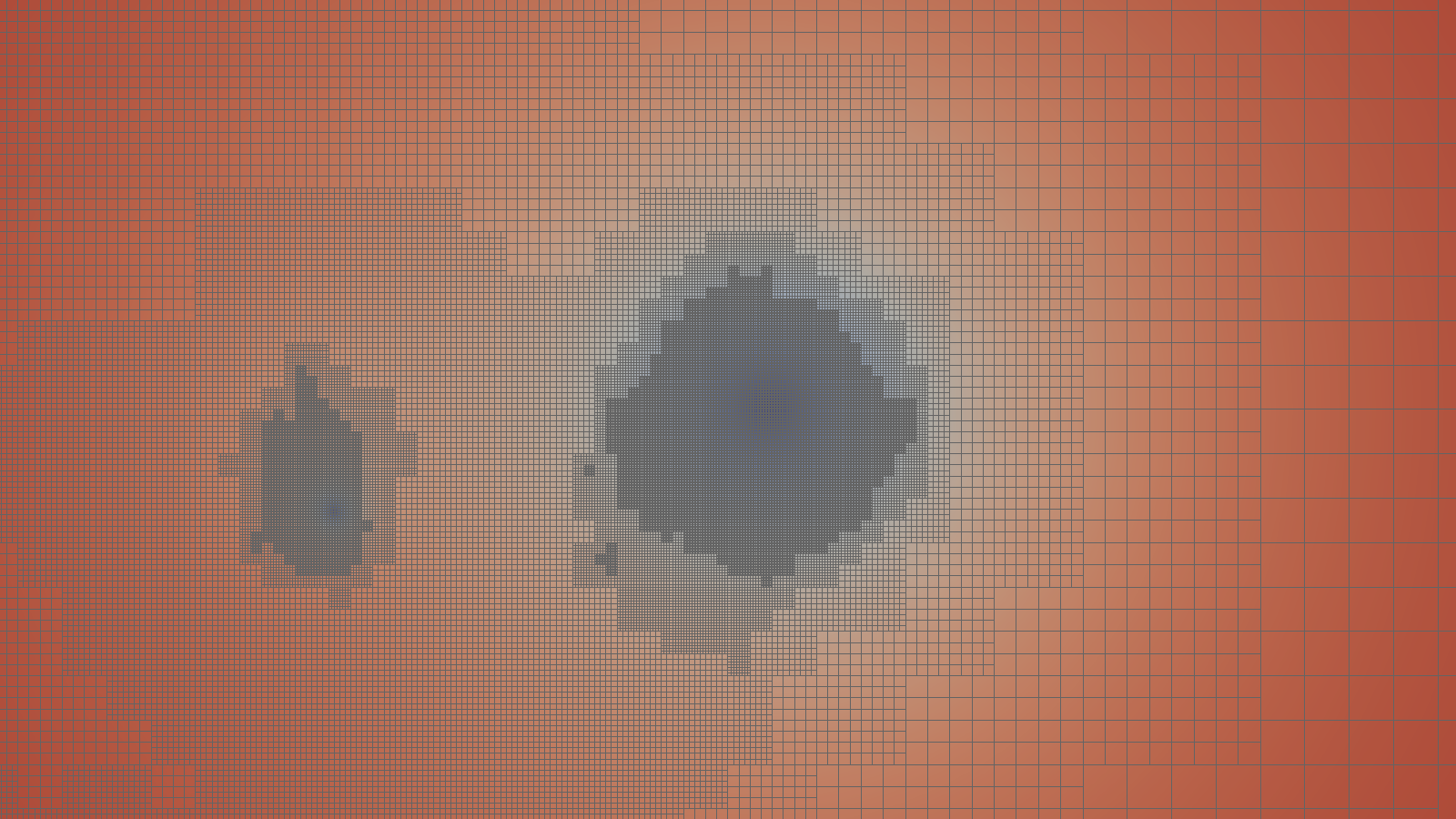}
	\includegraphics[width=0.32\columnwidth]{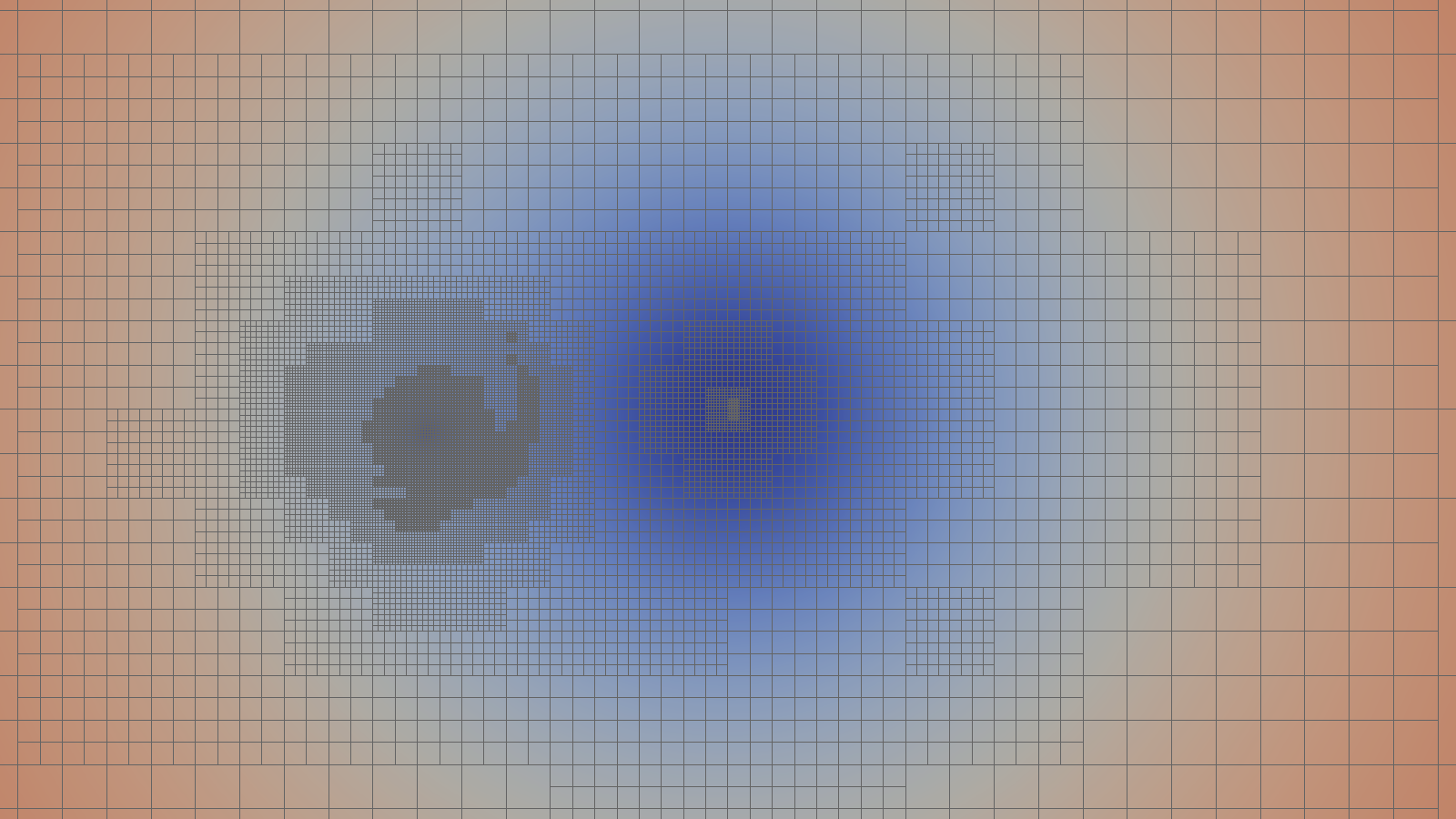}
	\caption{\small Example octree grids generated for black hole binaries with mass ratio 1,10,100. \label{fig:mr_bbh}}
\end{figure}

Adaptivity in spacetime can be vital for some computational applications, especially when spacetime adaptivity is necessary to make these simulations feasible even on leadership architectures. Here we present estimated speed up by using \nuts, for the simulation of binary black hole mergers and the computation of the resulting gravitational waves\cite{einsteinathome,LIGO,had_webpage,Fernando2018_GR}. The computational cost of these simulations increase significantly when the mass ratio $q$ of the two black holes increases. Assuming we need $n$ grid points in 1d to capture the larger black hole, to capture the smaller black hole in the presence of the larger black hole we need $qn$ in 1d, hence, increase of $q^3$ points in 3d. This makes the large mass ratio gravitational wave simulations infeasible at the time. We use these large mass ratio binary black hole grids(see figure \ref{fig:mr_bbh}) to estimate the speed up that can be enabled by the time adaptivity (see table \ref{tab:bbh}). For mass ratios of 10, one can expect up to \texttt{70x} speed up, which is a significant reduction in the cost (runtime and energy) of estimating the gravitational waves and can reduce the burden on supercomputing resources. 

\begin{table}[tbh]
    \centering
    \begin{tabular}{ |c|c|c|c|} 
        \hline
        mass ratio & $l_{min}$ & $l_{max}$ & est. speed up \\
        \hline
        1 &	3 &	15 & 9.82 \\
        2 &	3 &	16 & 18.105 \\
        5 &	3 &	18 & 15.035 \\
        10 &3 &	22 & 71.8302 \\
        \hline
    \end{tabular}
    \caption{Estimated speed up for binary black hole grid with increasing mass ratios from 1 to 10. }
    \label{tab:bbh}
\end{table}{}

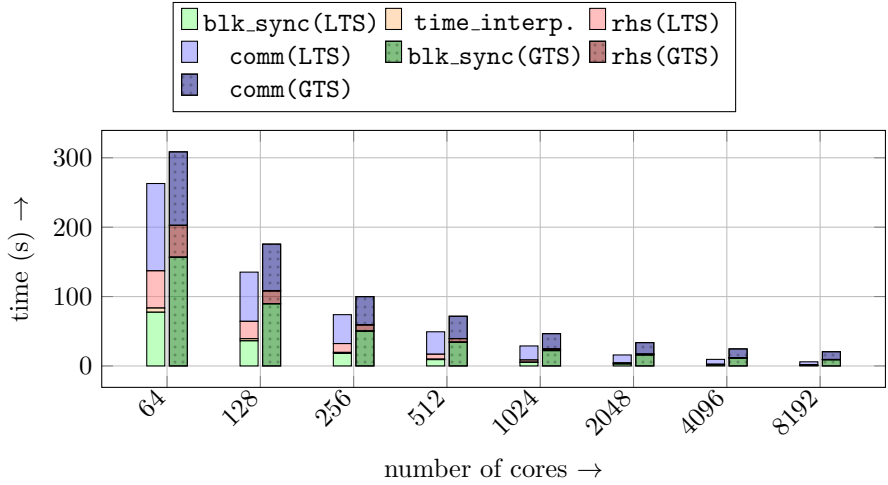
\begin{figure}[tbh]
	\centering
	\begin{tikzpicture}
	\begin{axis}[
	ybar stacked, bar width=0.24cm,    
	xlabel={number of cores $\rightarrow$},
	ylabel={time (s) $\rightarrow$ },symbolic x coords={64,128,256,512,1024,2048,4096,8192},width=12cm,height=5cm,
	xtick = data, 
	x tick label style={rotate=45, anchor=east, align=right},
	legend style={at={(0.45,1.5)}, anchor=north},legend columns=3,grid=major]
	\addplot [fill=green!50,  fill opacity=0.5] [bar shift=-.15cm] table[x={act_npes}, y = {blk_unzip}]{dat/enuts_ss1_frontera.dat};
	\addplot [fill=orange!50, fill opacity=0.5] [bar shift=-.15cm] table[x={act_npes}, y = {nuts_correct}]{dat/enuts_ss1_frontera.dat};
	\addplot [fill=red!50, fill opacity=0.5] [bar shift=-.15cm] table[x={act_npes}, y = {rhs}]{dat/enuts_ss1_frontera.dat};
	\addplot [fill=blue!50, fill opacity=0.5] [bar shift=-.15cm] table[x={act_npes}, y = {comm}]{dat/enuts_ss1_frontera.dat};
	\resetstackedplotsTwo
	\addplot [fill=green!50!black, fill opacity=0.5, postaction={pattern=dots}] [bar shift=.15cm] table[x={act_npes}, y = {unzip}]{dat/ets_ss1_frontera.dat};
	\addplot [fill=red!50!black  , fill opacity=0.5, postaction={pattern=dots}] [bar shift=.15cm] table[x={act_npes}, y = {rhs}]{dat/ets_ss1_frontera.dat};
	\addplot [fill=blue!50!black , fill opacity=0.5, postaction={pattern=dots}] [bar shift=.15cm] table[x={act_npes}, y = {comm}]{dat/ets_ss1_frontera.dat};
	\legend{\texttt{blk\_sync(LTS)},\texttt{time\_interp.}, \texttt{rhs(LTS)}, \texttt{comm(LTS)}, \texttt{blk\_sync(GTS)}, \texttt{rhs(GTS)}, \texttt{comm(GTS)}  }
	\end{axis}
	\end{tikzpicture}
	\caption{\small 
	Strong scaling results on TACC's \Frontera~ for \lts (left) and \gts (right) timestepping using the RK3 explicit scheme. For this experiment, the refinement levels are $l_{max}=10$ and $l_{min}=3$, hence $\Delta L=7$. Therefore a single \nuts~ is equivalent to $2^{\Delta L}=128$ global timesteps. For \lts, the plot shows the overall cost breakdown between block synchronization (\texttt{blk\_sync}), applying time interpolations between blocks (\texttt{LTS\_interp.}), computing the right hand side (\texttt{rhs}) and the communication costs(\texttt{comm}). For \gts, we show the cost breakdown between \texttt{blk\_sync}, \texttt{rhs} and \texttt{comm} (time interpolations are not required for \gts). The significant difference of \texttt{blk\_sync} cost between \lts~ and \gts. For \gts~ \texttt{blk\_sync} is a global operation, while in \lts~  \texttt{blk\_sync} is a local operation, where synchronization performed only on the subset of blocks, which are currently being evolved.Presented strong carried out for a fixed problem size of $262M$ unknowns where the number of cores ranging from $64$ to $8192$ cores. Note that for strong scaling results re-meshing is disabled in order to keep the problem size fixed and unchanged during evolution.
	\label{fig:ss_enuts_ss} 
	}
\end{figure}

\begin{figure}[tbh]
	\centering
\begin{tikzpicture}
	\begin{axis}[
	ybar stacked, bar width=0.4cm,    
	xlabel={number of cores $\rightarrow$},
	ylabel={time (s) $\rightarrow$ },symbolic x coords={8,16,32,64,128,256,512,1024,2048},width=10cm,height=5cm,
	xtick = data, 
	x tick label style={rotate=45, anchor=east, align=right},
	legend style={at={(0.5,1.3)}, anchor=north},legend columns=4,grid=major]
	\addplot [fill=orange!50, fill opacity=0.5] [bar shift=0cm] table[x={npes}, y = {t_rm_mean}]{dat/mesh_ws_frontera.dat};
	\addplot [fill=green!50, fill opacity=0.5] [bar shift=0cm] table[x={npes}, y = {t_bal_mean}]{dat/mesh_ws_frontera.dat};
	\addplot [fill=red!50, fill opacity=0.5] [bar shift=0cm] table[x={npes}, y = {t_mesh_mean}]{dat/mesh_ws_frontera.dat};
    \legend{\texttt{SFC\_Wpart}, \texttt{bal.(2:1)}, \texttt{mesh\_gen}}
	\end{axis}
\end{tikzpicture}
\caption{\small Weak scaling results in TACC's \Frontera~ to perform SFC-based weighted partitioning (\texttt{SFC\_Wpart}) for Gaussian octant distribution centered at (0,0,0) followed by $2:1$ balancing (\texttt{bal.2:1}) of the octants, which is used as the input for the mesh generation(\texttt{mesh\_gen}). For this experiment, we used 1.6M grid points per core, using $7^3$ points per octant, with the number of cores varying from $8$ to $2048$. The largest problem reported had a total of $3.3B$ grid points, where the mesh generation completed under 2s.}  
\label{fig:mesh_ws}
\end{figure}
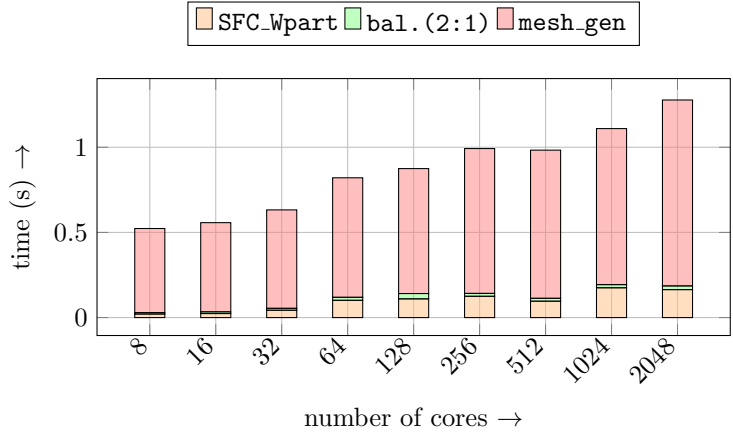

\begin{figure}[tbh]
	\centering
\begin{tikzpicture}
	\begin{axis}[
	ybar stacked, bar width=0.4cm,    
	xlabel={number of cores $\rightarrow$},
	ylabel={time (s) $\rightarrow$ },symbolic x coords={32,64,128,256,512,1024,2048},width=10cm,height=5cm,
	xtick = data, 
	x tick label style={rotate=45, anchor=east, align=right},
	legend style={at={(0.5,1.3)}, anchor=north},legend columns=4,grid=major]
	\addplot [fill=orange!50, fill opacity=0.5] [bar shift=0cm] table[x={npes}, y = {t_rm_mean}]{dat/mesh_ss_frontera.dat};
	\addplot [fill=green!50, fill opacity=0.5] [bar shift=0cm] table[x={npes}, y = {t_bal_mean}]{dat/mesh_ss_frontera.dat};
	\addplot [fill=red!50, fill opacity=0.5] [bar shift=0cm] table[x={npes}, y = {t_mesh_mean}]{dat/mesh_ss_frontera.dat};
    \legend{\texttt{SFC\_Wpart}, \texttt{bal.(2:1)}, \texttt{mesh\_gen}}
	\end{axis}
\end{tikzpicture}
\caption{\small Strong scaling results in TACC's \Frontera~ to perform SFC-based weighted partitioning (\texttt{SFC\_Wpart}) for Gaussian octant distribution centered at (0,0,0) followed by $2:1$ balancing (\texttt{bal.2:1}) of the octants, which is used as the input for the mesh generation(\texttt{mesh\_gen}). For the depicted strong scaling, we keep the problem size fixed at $3.3B$ grid points with the number of cores increasing from $32$ to $2048$.}  
\label{fig:mesh_ss}
\end{figure}
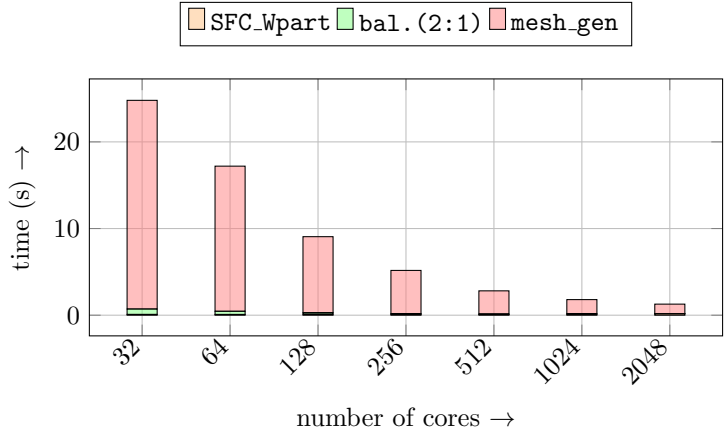

\subsection{Weak and strong scaling}

Parallel scalability of timesteppers is essential when dealing with large scale simulations. In this section, we present weak and strong scalability results for the linear wave propagation problem, on octree meshes using global and local timestepping. For both weak and strong scaling, we used $7^3$ grid points per octant. 
For weak scaling, we set the computational domain to $[-10,10]^3$, and used \maxDepth~ 10. For the \maxDepth~ 10 grid generated $l_{min}=3$ and $l_{max}=8$, hence 32 global timesteps is equivalent to a single \lts~ timestep. Therefore, in order to make the \uts~ and \nuts~ results comparable, in the following scaling results, we present timing for 32 steps in \gts, 1 step (32 partial steps) in \lts. In weak scaling, we increase the grid size, such that keeping the degrees of freedom per core roughly constant (100K unknowns per core).  Weak scaling results for \lts~ and \gts~ are presented in figure \ref{fig:ws_enuts_g300}.  Each bar presents the corresponding evolution time between \lts~ and \gts. For each scheme, we present the overall cost breakdown between, computation of the right-hand side(\texttt{rhs}) of the PDE, and block padding synchronization ($\texttt{blk\_sync}$) and inter-process communication (\textit{comm}). The \texttt{blk\_sync} cost consists of space interpolation, which is common for both \lts~ and \gts~ schemes due to space adaptivity. We present an extra fraction of time interpolation cost between blocks only present in the \lts. For \gts~ the \texttt{blk\_sync} operation is a global synchronization, i.e., all the blocks are evolved and need to synchronize the padding regions for the next \texttt{rhs} computation. In contrast, \lts~ scheme the \texttt{blk\_sync} operation is a local(partial) synchronization limited to the blocks evolved at the current partial step followed by time interpolation to correct the padding regions between blocks. The weak scaling plot shows that the partial synchronization with appropriate time interpolation is efficient than the global synchronization in \gts~ scheme.

To perform strong scaling (see figure \ref{fig:ss_enuts_ss} ), we use the \maxDepth~ 12 and recorded $l_{min}$ and $l_{max}$ were 3 and 10, respectively. Therefore, 128 global timesteps are equivalent to a single \lts~ timestep. For strong scaling tests, we keep the total grid size fixed $262M$ unknowns and increase the number of cores from $64$ to $8192$. The strong scaling plot shows, the same cost breakdown described above. The recorded average parallel efficiencies between \lts~ and \gts~ schemes were $87\%$ and $74\%$, respectively. The low overhead of \texttt{blk\_sync} operation, allows \lts~ to demonstrate superior weak and strong scalability compared to \gts~ scheme.

\subsection{Weighted partitioning and mesh-generation}
The performance of \\ data partitioning and mesh generation is crucial for AMR applications, especially when the computational grid changes frequently. We refer to this process as re-meshing, which require re-partitioning of the data (since the refinement change may have caused load-imbalance), enforcing 2:1 balancing, and mesh data structure generation. The performance of the re-meshing is crucial, but it is not the main focus of this paper. In the current implementation, we trigger refinement in \lts~ when all the blocks are synchronized in time. Figure \ref{fig:lsm} shows how the grid changes as the wave propagate radially outwards. Figures \ref{fig:mesh_ws} and \ref{fig:mesh_ss} show the weak and strong scalability of the operations related to re-meshing. The above experiments show that mesh generation has a relatively high computational cost, compared to SFC weighted partitioning and 2:1 balancing of octrees. This is mainly because mesh generation performs a large number of search operations on the octree to build the neighborhood data structures, which are essential to perform numerical computations.

\section{Conclusions}
\label{sec:conclusion}
In this paper, we presented methods to enable time adaptivity for solving PDEs numerically on spatially adaptive grids. We presented experimental results for the accuracy and scalability of the presented approaches. We show that for some highly adaptive octrees with high levels of refinement, time adaptivity can be essential to reduce overall time to solution. As future work, we will explore more sophisticated energy-conserving time interpolation operators, with increased accuracy. Currently, re-partitioning, and re-meshing are triggered as global operations when all the blocks are synchronized in time. If only a localized portion of the mesh needs to be refined, then ideally one should only require an update (remesh) of those regions. This is complicated by the fact that different regions will no longer be syncronized in time. While the proposed \lts\ scheme supports this, performing this efficiently requires new partial remeshing algorithms, that will be the focus of future work.


\bibliographystyle{siamplain}
\bibliography{bssn,nuts}


\end{document}